\documentclass[aps,pra,floatfix,twocolumn,footinbib,amsmath,amssymb]{revtex4-1}
\usepackage                             {epsfig}
\usepackage                             {natbib}
\usepackage                             {xspace}
\usepackage                             {amsmath}
\usepackage                             {graphicx}
\usepackage     [english]               {babel}
\usepackage                             {natbib}
\usepackage                             {hyperref}
\usepackage                             {amsfonts}
\usepackage                             {amssymb}
\usepackage                             {latexsym}
\usepackage                             {color}
\newcommand{\figref}[1]{Fig.~\ref{#1}}

\newcommand{\secref}[1]{Section~\ref{#1}}
\newcommand{\eref}[1]{(\ref{#1})}
\renewcommand{\eqref}[1]{Eq.~(\ref{#1})}
\newcommand{\tabref}[1]{Table~\ref{#1}}
\newcommand{\ket}[1]{| #1 \rangle}

\newcommand{\braket}[1]{\langle #1 \rangle}
\begin{document}
\title{Theory of inelastic confinement-induced resonances due to the
  coupling of center-of-mass and relative motion}

\author{Simon Sala and Alejandro Saenz}

\affiliation{AG Moderne Optik, Institut f\"ur Physik,
  Humboldt-Universit\"at zu Berlin, Newtonstrasse 15, 12489 Berlin,
  Germany}
\date{\today}
\begin{abstract}
  A detailed study of the anharmonicity-induced resonances caused by
  the coupling of center-of-mass and relative motion is presented for
  a system of two ultracold atoms in single-well potentials.  As has
  been confirmed experimentally, these inelastic confinement-induced
  resonances are of interest, since they can lead to coherent molecule
  formation, losses, and heating in ultracold atomic gases. A
  perturbative model is introduced to describe the resonance positions
  and the coupling strengths. The validity of the model and the
  behavior of the resonances for different confinement geometries are
  analyzed in comparison with exact numerical \textit{ab initio}
  calculations.  While such resonances have so far only been detected
  for large positive values of the $s$-wave scattering length, it is
  found that they are present also for negative $s$-wave scattering
  lengths, i.\,e.\ for attractive interactions. The possibility to
  coherently tune the resonances by a variation of the external
  confinement geometry might pave the way for coherent molecule
  association where magnetic Feshbach resonances are inaccessible.
\end{abstract}
\maketitle
%
%
\section{Introduction}

Theoretical treatments of strongly-correlated ultracold atoms in
single-well potentials routinely adopt the harmonic approximation to
describe the trapping potential. This is, of course, an idealization,
because every realistic trapping potential is finite and thus
anharmonic. A major benefit of a harmonic confinement is that for
identical particles relative (rel.)\ and center-of-mass (c.m.)\ motion
decouple. Moreover, especially for deep potentials the harmonic
confinement resulting from a second-order Taylor expansion of the
potential in its minimum might be a good approximation, especially
since the center-of-mass to relative motion (c.m.-rel.)\ coupling
introduced by the anharmonicity of the trapping potential is
energetically negligible compared to the energy scale of the
interatomic interaction.  As a consequence, the theoretically
predicted binding energy of two ultracold atoms confined in a harmonic
trap \cite{cold:busc98} has been experimentally confirmed
\cite{cold:stoe06}. Moreover, the measurement \cite{cold:deur08} and
calculation \cite{cold:gris09} of the influence of the anharmonicity
on the energy of states in deep optical lattices has revealed that the
deviation to the harmonic approximation for the lowest band is
negligible in most cases.

On the first glance, it was therefore surprising that c.m.-rel.\
coupling can have a significant impact on an ultracold atomic quantum
gas. In \cite{cold:sala12,cold:peng11} it was revealed that the
particle loss and heating observed in \cite{cold:hall10b} was caused
by inelastic confinement-induced resonances (CIR), i.\,e.\ resonances
due to the c.m.-rel.\ coupling of a c.m.\ excited bound state with an
unbound atom pair -- even in a deep optical lattice where the harmonic
approximation has proven accurate for the rel.\ motion energies. The
explanation of the losses due to {\em inelastic} CIR was necessary in order
to resolve contradicting results of the experiment \cite{cold:hall10b}
and the theory of {\em elastic} CIR \cite{cold:olsh98,cold:berg03} that 
originally was used as an explanation for the observed atom losses. Additionally,
alternative explanations of the experimental findings were proposed
based on the assumption of a harmonic confinement. One is based on
multichannel effects \cite{cold:mele11}, others \cite{cold:hall10b,
  cold:arim11} on a Feshbach-type mechanism. Finally, in
\cite{cold:sala13} it was demonstrated by simultaneously 
observing both the elastic and the inelastic CIRs 
and by excluding other mechanisms like many-body effects 
due to the design of the experiment that a coherent molecule formation is
triggered uniquely at a c.m.-rel.\ coupling resonance which confirms
the explanation of the losses in \cite{cold:hall10b} in terms of the 
inelastic CIR. This
demonstrates the importance of the anharmonicity-induced c.m.-rel.\
coupling, especially in view of its universality.

In fact, very recently it was demonstrated that inelastic CIR are also present
in ultracold dipolar systems \cite{cold:schu15}, and even for
Coulomb-interacting systems such as excitons in quantum-dot systems 
\cite{cold:trop15}. For the latter, the resonances were proposed to serve 
as a novel kind of controlled single-photon source. At the inelastic CIR a
variation of the exciton confinement leads to a redistribution of the
charge density with subsequently increased annihilation probability of
the electron-hole pair. Since this process can be steered {\it in situ} 
by a variation of the external confinement, single photons can
be emitted on demand.

Resonances in atomic gases due to c.m.-rel.\ coupling have been
mentioned in literature before. The first work explicitly discussing a
possible molecule formation due to anharmonicity-induced c.m.-rel.\
coupling is \cite{cold:bold05}. In the model in \cite{cold:bold05} a
deep isotropic optical lattice is considered. The evaluation of the
c.m.-rel.\ coupling matrix elements is performed with wavefunctions in
the harmonic approximation. A direct loss process is considered where
two unbound atoms couple to a molecule in the continuum, i.\,e.\ the
c.m.-part of the bound state is assumed to be a highly excited c.m.\
wavefunction which is approximated by a spherical wave. The work
concludes that the dimer-production rate is too small to be
significant in an optical-lattice experiment. The observed losses in
\cite{cold:hall10b} proof different, however, i.\,e.\ losses at
inelastic CIR in a deep optical lattice are observed. The reason for
the small dimer-production rate in \cite{cold:bold05} is that the
coupling to highly excited bound states is very small as will be
 shown in this work. On the other hand, as demonstrated in
\cite{cold:sala12, cold:sala13}, losses at inelastic CIR occur
dominantly due to the coupling of a rel.-motion bound state with 
low-order c.m.\ excitation and subsequent three-body collisions, 
in contrast to a direct coupling to a rel.-motion bound state 
in the c.m.\ continuum as assumed in
\cite{cold:bold05}.
 
Even in a harmonic confinement a coupling of the c.m.\ and rel.\
motion is present in the case of heteronuclear atoms or identical 
atoms in different electronic 
states \cite{cold:pean05,cold:gris07,cold:mele09}. Moreover, the occurrence of
Feshbach-type resonances due to the c.m.-rel.\ coupling was discussed
in \cite{cold:schn09}, their behavior in a superlattice was
characterized in \cite{cold:kest10}. In mixed dimensions, the
experiment performed in \cite{cold:lamp10} also detected inelastic
loss resonances for a variation of the scattering length. The behavior
of the elastic and inelastic CIR within a quasi-1D lattice model was
considered in \cite{cold:vali11}.

In the present work, the two-body model that was introduced in
\cite{cold:sala12} for the description of the positions of the resonances in
quasi-1D and quasi-2D confinement observed in \cite{cold:hall10b} is
generalized to describe the position and coupling strength of in
principle all c.m.-rel.\ coupling resonances in single-well potentials
of arbitrary symmetry.  The validity of the model is discussed in
comparison to \textit{ab initio} calculations. Resonance positions and
coupling strengths are investigated for a change in the confinement
geometry, i.\,e.\ in the transition between an almost isotropic to a
cigar-shaped (quasi-1D) confinement and in the transition between an
almost isotropic to a pancake-shaped (quasi-2D) confinement. It is
demonstrated that higher-order resonances occur also for negative
values of the $s$-wave scattering length. Consequently, by analyzing
the wavefunction of the system it will be demonstrated that molecule
formation also occurs in the strongly attractive interaction regime.

The paper is organized as follows: In \secref{sec:hamil} the two-body
Hamiltonian is introduced and the basic idea of inelastic CIR is
briefly recapitulated. Then a model is formulated to predict the
position (\secref{sec:position}) and coupling strength
(\secref{sec:coupling}) of c.m.-rel.\ resonances. To validate the
model, the behavior of the resonances is discussed under a varying
geometry of the confining potential in Sections
\ref{sec:transition_1d} and \ref{sec:transition_2d}. \secref{sec:opt}
deals with the optimization of c.m.-rel.\ coupling and the limitations
of the model. It is then concluded in \secref{sec:wf} that resonances
ignored in previous considerations (e.\,g.\ in \cite{cold:sala12})
should also lead to a molecule formation in the strongly attractive
interaction regime of ultracold atoms. The article closes with a
conclusion in \secref{sec:conclusion}.

\section{The Hamiltonian}
\label{sec:hamil}
\begin{figure}[ht]
  \begin{centering}
    \includegraphics[width=0.40\textwidth]{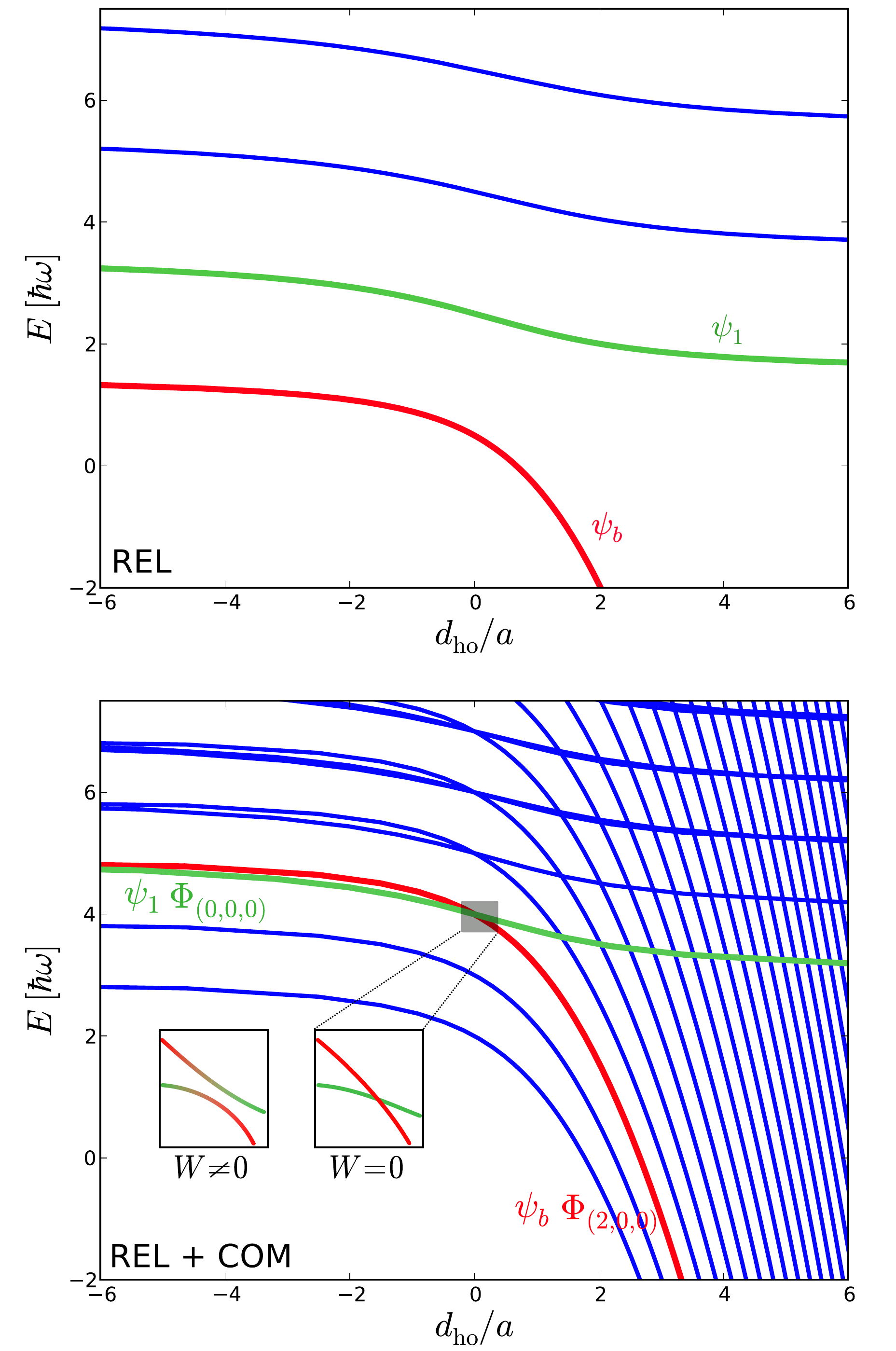}
    \includegraphics[width=0.38\textwidth]{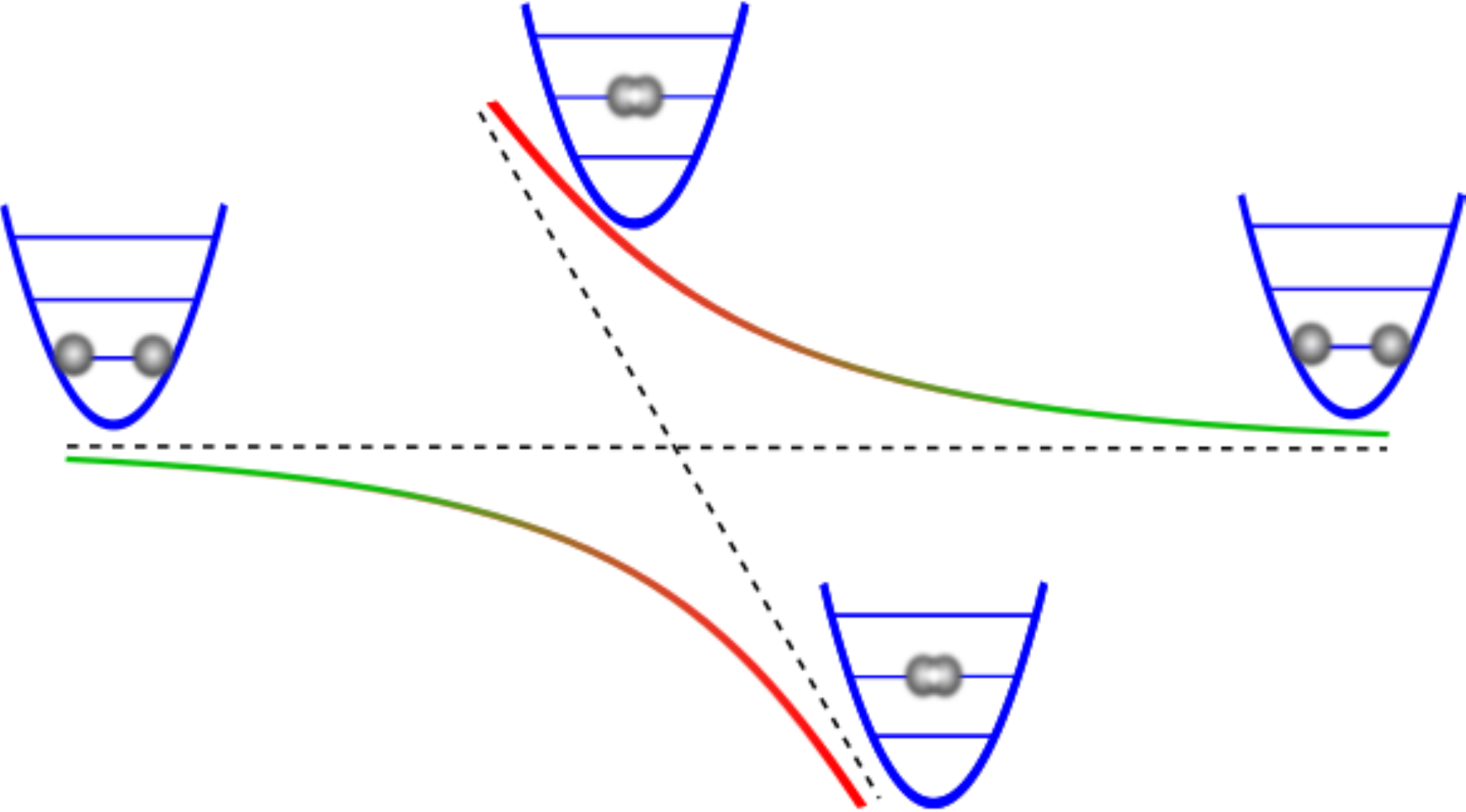}
    \caption{(Color online) Eigenenergy spectrum of two ultracold atoms
    in an isotropic harmonic trap interacting via a $\delta$
    pseudopotential \cite{cold:busc98, cold:idzi06} for a varying
    $s$-wave scattering length $a$ (or varying confinement length
    $d_{\rm ho}$ which would, however, also change $\omega$
    continuously). The upper panel shows the energy of the rel.\
    motion Hamiltonian $H_{\mathrm{rel}}$, the lower one the combined energy
    of the rel.\ and the c.m.\ motion Hamiltonians, i.\,e.\ $H_{\mathrm{rel}} +
    H_{\mathrm{cm}}  $. Introducing a coupling between the rel.\ and
    c.m.\ motion makes the crossings avoided as illustrated in the
    lower part as solid lines, while the black dashed lines indicate the
    diabatic curves.  Passing through the crossing adiabatically (on
    the solid line) allows for a transformation of the bound state
    with c.m.\ excitation into a trap state in the c.m.\ ground
    state. }
  \label{fig:basic_spec}
  \end{centering}
\end{figure}

In c.m.\ and rel.\ coordinates, $\mathbf{r}=
\mathbf{r_1}-\mathbf{r_2}$ and
$\mathbf{R}=\frac{1}{2}(\mathbf{r_1}+\mathbf{r_2})$, respectively, the
Hamiltonian of two identical particles in an external trapping
potential can be written as
\begin{align}
\label{eq:Hamil}
  & H(\mathbf{r},\mathbf{R}) =\ H_{\mathrm{rel}}(\mathbf{r}) + H_{\mathrm{cm}}(\mathbf{R}) + W(\mathbf{r},\mathbf{R}) \\
\label{eq:Hamil_rel}
  & H_{\mathrm{rel}}(\mathbf{r})  =\ T_{\rm rel}(\mathbf{r}) + V_{\rm rel}(\mathbf{r}) + U_{\rm int}(r)\\ 
\label{eq:Hamil_com}
  & H_{\mathrm{cm}}(\mathbf{R}) =\  T_{\rm cm}(\mathbf{R}) + V_{\rm cm}(\mathbf{R}) \quad.
\end{align}%
where $T_{\mathrm{rel}}$ and $T_{\mathrm{cm}}$ denote the
kinetic-energy operators of the rel.\ and c.m.\ motion, respectively,
and $V_{\rm rel}$ and $V_{\rm cm}$ are the separable parts of the
potential energy. Thus, $W$ contains only the non-separable terms of
the potential energy. $U_{\rm int}(r)$ is the interparticle
interaction.

In \figref{fig:basic_spec} the eigenenergy spectrum of the Hamiltonian
in \eqref{eq:Hamil} for two ultracold atoms interacting via the
$\delta$ pseudopotential
\begin{align}
  U_{\rm int}(r)=\frac{4 \pi \hbar^2 a}{m}
  \delta(\mathbf{{r}})\frac{\partial}{\partial r}r \quad ,
\end{align}
$a$ being the $s$-wave scattering length and $m$ the atomic mass, is
shown. The atoms are confined in an isotropic, i.\,e.\ spherically
symmetric, harmonic trapping potential.
The energies are plotted for a varying inverse $s$-wave scattering
length $d_{\mathrm{ho}}/a$, $d_{\mathrm{ho}} = \sqrt{\frac{2 \hbar}{m
    \omega}}$ being the harmonic oscillator length.
The spectrum of $H_{\mathrm{rel}}$ (\eqref{eq:Hamil_rel}) contains a
bound state $\ket{\psi_b}$ bending down to negative infinity for $a
\to 0^+$ and trap states, the energetically lowest one denoted as
$\ket{\psi_1}$. In case of a harmonic trapping potential, the coupling
$W$ between the c.m.\ and the rel.\ motion vanishes. Hence, in order
to obtain the spectrum of the full Hamiltonian (\eqref{eq:Hamil}) the
c.m.\ energies are added to each state of the rel.\ spectrum,
resulting in the middle plot of \figref{fig:basic_spec}. Each state of
the rel.\ spectrum appears now with an infinite series of c.m.\
excitations. Crossings appear between c.m.\ excited bound states,
e.\,g., $\ket{\psi_b \Phi_{(2,0,0)}}$, and trap states, e.\,g., the
lowest trap state $\ket{\psi_1 \Phi_{(0,0,0)}}$. Introducing a
coupling $W(\mathbf{r},\mathbf{R}) \neq 0$ between c.m.\ and rel.\
motion turns the crossings to avoided ones (besides modifying the
energies of the states also globally) and enables an adiabatic
transition of a c.m.-excited molecular state and a state of an unbound
atom pair in the c.m.\ ground state a indicated in the lower plot of
\figref{fig:basic_spec}.

Therefore, the c.m.-rel.\ coupling introduces a Feshbach-type
resonance at the crossing position. The occupation of the bound state
at the resonance is only possible because the excess binding energy
can be transferred into c.m.\ excitation energy due to the
anharmonicity of the confining potential. This redistribution of
binding energy to kinetic energy marks an inelastic process and thus
these c.m.-rel.\ coupling resonances were denoted as \textit{inelastic}
CIR. In \cite{cold:sala13} it is demonstrated how a coherent molecule
formation can be realized at the c.m.-rel.\ coupling resonance. Such a
molecule formation can trigger particle loss and heating in a
many-body system in a two-step process. At the resonance two atoms
coherently couple to the c.m.-excited molecular state. Then, this
molecule collides either with another molecule or an unbound atom
leading to a deexcitation of the molecule into a deeply bound state and
subsequent loss of the involved particles from the trap.

The full two-body spectrum \figref{fig:basic_spec} shows a plethora of
crossings. A full six-dimensional treatment of the two-body problem
involving the c.m.-rel.\ coupling is possible \cite{cold:gris09,cold:gris11}, but 
numerically quite demanding.  Since c.m.-rel.\ coupling resonances can have a
substantial influence on the stability of an ultracold atomic gas, the
knowledge of the position and coupling strength of the resonances is
of great interest. The understanding and assignment of these resonances is in
particularly important for an unbiased identification of other few-body 
resonances of, e.\,g., Efimov type. Therefore, simplified models that 
allow for the estimate of resonance positions and widths are desirable.  

\section{C.m.-rel.\ coupling model -- resonance positions}
\label{sec:position}

A model is introduced to predict the position and coupling strength of
the inelastic CIR. The theory generalizes the model described in
\cite{cold:sala12} that is only applicable to strongly anisotropic
(quasi-1D or quasi-2D) confinement and does not consider coupling
strengths explicitly \footnote{In more detail: Compared to the model
  presented in \cite{cold:sala12}, the here presented model uses a
  different formula for the bound-state energy and the anharmonic
  corrections are treated differently. These changes extent
  substantially the applicability of the model. The coupling strengths
  were not treated explicitly in previous works.}.

Of course, for c.m.-rel.\ coupling resonances to be present a
c.m.-rel.\ coupling must be introduced. Therefore, the harmonic
approximation has to be abandoned. Optical-lattice potentials
\cite{cold:bloc05} are widely used in ultracold experiments and offer
a great degree of flexibility and control. In a deep optical lattice,
i.\,e.\ $\frac{E_r}{V} \ll 1$ where $E_r = \frac{\hbar^2 k^2}{2 m }$ is
the recoil energy and $V$ is the lattice depth, tunneling between
neighboring wells is suppressed and the potential can be regarded as a
stack of single-well potentials. In this work, a sextic potential is
used resulting from a Taylor expansion of a $\sin^2$ optical-lattice
potential up to the sixth degree. A separation of the expansion in
rel., c.m., and coupling terms
\begin{align}
  \label{eq:sextic_rel}
  &V_{\mathrm{rel}}(\mathbf{r}) = \sum_{j=x,y,z}V_j \left[ \frac{1}{2} k_j^2 r_j^2 - 
                      \frac{1}{24}k_j^4r_j^4 +
                       \frac{1}{720}k_j^6r_j^6 \right] \\
  \label{eq:sextic_com}
  &V_{\mathrm{cm}}(\mathbf{R}) =  \sum_{j=x,y,z}V_j \left[2 k_j^2 R_j^2 - 
                      \frac{2}{3} k_j^4R_j^4 +
                       \frac{4}{45} k_j^6R_j^6 \right] \\
  \label{eq:sextic_w}
  &W(\mathbf{r}, \mathbf{R}) =  \sum_{j=x,y,z}  V_j \left[ -k_j^4 r_j^2 R_j^2
    +\frac{1}{3}k_j^6 r_j^2 R_j^4 
    + \frac{1}{12}k_j^6 r_j^4 R_j^2 \right],
\end{align}
respectively, shows that the quartic terms all have a negative sign
which makes the expansion to the sextic degree necessary. Otherwise, an  
unphysical continuum would occur in the spectrum reaching in energy
towards negative infinity. In Eqs.\
\eref{eq:sextic_rel}-\eref{eq:sextic_w} $V_j$ is the lattice depth in
direction $j\in\{x,y,z\}$, $k_j=\frac{2\pi}{\lambda_j}$, and
$\lambda_j$ is the wavelength of the laser in direction
$j$. Introducing the harmonic oscillator frequencies $\omega_j^2 =
\frac{2V_j}{m}k_j^2$ the potential terms can be written in a more
canonical form.

It has been demonstrated \cite{cold:gris09,cold:sala12,cold:sala13}
that sextic potentials are well suited to describe
anharmonicity-induced c.m.-rel.\ coupling in single-well
potentials. The large flexibility in the potential parameters makes
the sextic potential, moreover, suitable to accurately describe the
c.m.-rel.\ coupling in a variety of potentials, e.\,g.\ Gaussian beam
potentials \cite{cold:sala13}.

It is first assumed that the anharmonicity only has a small (global)
influence on the eigenenergies of the states. Hence, the harmonic
approximation is used and the position of a c.m.-rel.\ coupling
resonance is determined by the position of the crossing of the c.m.\
excited bound state $\ket{\psi_b \Phi_{\mathbf{n}}}$, where
$\mathbf{n} = (n_x,n_y,n_z)$ is the quantum number of the c.m.\
excitation, that separates spatially for an optical-lattice potential,
and a trap state $\ket{\psi_t \Phi_{\mathbf{m}}}$ as illustrated in
\figref{fig:basic_spec}.

The energy $E_b^{\mathrm{rel}}$ of the bound state $\ket{\psi_b}$ in
an harmonic confinement of arbitrary anisotropy in dependence of the
$s$-wave scattering length $a$ is given implicitly by \cite{cold:lian08}  
\begin{align}
   \label{eq:bound_state_ener} 
   &\frac{\sqrt{\pi} d_y}{a} = \nonumber  \\
   &-\int_{0}^{\infty } \left( \frac{\sqrt {\eta_x \eta_z}{{\rm
           e}^{\frac{t\epsilon}{2}}}} {\sqrt { \left( 1-{{\rm e}^{-t}}
         \right) \left( 1-{{\rm e}^{-\eta_x\,t}} \right) \left(
           1-{{\rm e}^{-\eta_z\,t}} \right) }} -t^{-\frac{3}{2}}
   \right)dt
\end{align}
where $\epsilon=(E_b^{\rm rel}-E_0)/(\hbar \omega_y)$,
$E_0=\frac{\hbar}{2}(\omega_x + \omega_y + \omega_z)$,
$\eta_x=\omega_x/\omega_y$, and $\eta_z=\omega_z/\omega_y$. 
The difference to other equations for the bound state, like e.\,g.\ in
\cite{cold:busc98} (valid for a 3D isotropic confinement), in
\cite{cold:idzi06} (valid for a 3D harmonic confinement of single
anisotropy), or in \cite{cold:peng10} (valid for only transversally
trapped atoms), is that it is valid for an arbitrarily anisotropic 3D
confinement.

A general expression for the eigenenergies of the trap states in an
arbitrarily anisotropic confinement, i.\,e.\ states above $E_{\rm
  th}=\frac{\hbar}{2}(\omega_x + \omega_y +\omega_z)$, is not known
yet. For ultracold temperatures the occupation of excited states is
suppressed. Hence, in the following only crossings with the first trap
state $\ket{\psi_1}$ are considered. Assuming without loss of
generality that $\min(\omega_x,\omega_y,\omega_z) = \omega_z$ (unless
stated differently), the eigenenergy $E_{\rm 1}^{\rm rel}$ of
$\ket{\psi_1}$ lies in the interval $[E_{\rm th},E_{\rm th}+ 2 \hbar
\omega_z)$. It can be shown that in the case of an isotropic harmonic
confinement the crossing between an excited bound state with a single
lowest-order c.m.\ excitation ($n_i=2, n_{j \ne i}=0$) with the first
trap state occurs at
\begin{align}
  E_{\rm 1}^{\rm rel} = E_{\rm th}+\hbar \omega_z 
   \label{eq:1st_trap_ho}
\end{align}
which is thus chosen for the model as an approximation of the energy
of the first trap state.  For crossings with higher trap states, the
model can be extended by the proper choice of the energy of that trap
state.

For a spatially decoupled confinement, like expansions of an
optical-lattice potential, the eigenstates of the c.m.\ Hamiltonian
factorize as $\Phi_\mathbf{n}(\mathbf{R}) = {\phi_{n_x}({X})}\,
{\phi_{n_y}({Y})}\, {\phi_{n_z}({Z})}$ with eigenenergies
$E_\mathbf{n}^{\rm cm}=\sum_{j=x,y,z} \hbar \omega_j
(n_j+\frac{1}{2}$). When combining rel.\ and c.m.\ motions the
energies of the bound states become $E_b^{\rm
  rel}(a)+E_\mathbf{n}^{\rm cm}$ while the energy of the lowest trap
state is given by $ E_{\rm 1}^{\rm rel}+E_{(0,0,0)}^{\rm
  cm}$. Crossings between a c.m.\ excited bound state and the lowest
trap state are determined by
\begin{align}
  \label{eq:crossing} 
  E_b^{\rm rel}=E_{\rm 1}^{\rm rel}-\Delta_\mathbf{n}
\end{align} 
where 
\begin{align}
  \Delta_\mathbf{n}=E_\mathbf{n}^{\rm cm}-E_{(0,0,0)}^{\rm cm}
  \label{eq:Delta_ho}
\end{align}
is the c.m.\ excitation. The corresponding $s$-wave scattering length
at the crossing is obtained from \eqref{eq:bound_state_ener}.

So far all energies were treated within the harmonic approximation. It
will be demonstrated that this purely harmonic model gives in some
cases already good quantitative results. However, for the c.m.\
excitations $\Delta_{\mathbf{n}}$ higher c.m.\ states are
involved. Additionally, for small $\Delta_{\mathbf{n}}$ the bound
state crosses the trap state for small positive $d_{\mathrm{ho}}/a$ or
even negative $d_{\mathrm{ho}}/a$ where the two states cross with
comparable slopes, see \figref{fig:basic_spec}. For such crossings the
position is very sensitive to the energies of the involved
states. Hence, the energy of the first trap state as well as the c.m.\
excitation $\Delta$ must be corrected.

Treating the anharmonic terms of the c.m.\ potential in
\eqref{eq:sextic_com}, $\sum_{j=x,y,z} -\frac{1}{24}\frac{\hbar
  \omega_j}{V_j}R_j^4 + \frac{1}{720}\frac{\hbar^2
  \omega_j^2}{V_j^2}R_j^6$ (here written in dimensionless units of
energies in $\hbar \omega_j$ and lengths in $\sqrt{\frac{\hbar}{M
    \omega_j}}$ with $M=2m$) within first-order perturbation theory
results in 
\begin{align}
  \Delta_{(n_x,n_y,n_z)} = \sum_{j=x,y,z} & {\hbar \omega_j} \bigg[
    {n_j} - \frac{{\hbar \omega_j}}{16 V_j} \, {\left({n_j}^{2} +
        {n_j}\right)} \nonumber \\
    &+ \frac{{\hbar^{2} \omega_j^{2}}}{576 V_j^2} {\left(2 \,
        {n_j}^{3} + 3 \, {n_j}^{2} + 4 \, {n_j}\right)} \bigg]
\label{eq:Delta}
\end{align}
for the c.m.\ excitation and
\begin{align}
  E_{\rm 1}^{\rm rel} = \hbar \omega_z +
  \sum_{j=x,y,z} \frac{1}{2} \, {\hbar \omega_j} - \frac{{\hbar
      \omega_j}^{2}}{32 \, V_j} + \frac{{\hbar \omega_j}^{3}}{384 \,
    V_j^{2}}
\label{eq:1st_trap}
\end{align}
for the energy of the first trap state (see \secref{sec:perturb} in
the Appendix for details).

It will be demonstrated that in the case of resonances for negative
values of the $s$-wave scattering length and strongly anisotropic confinement,
the effective 1D c.m.\ problem needs even to be solved exactly and thus 
numerically to obtain accurate results. For the numerical evaluation
of the stationary 1D Schr\"odinger equation, the approach described in
\cite{cold:foer12} was used.

Therefore, three models for the resonance position were introduced
that differ by the treatment of the c.m.\ excitation $\Delta$ and by
the energy of the first trap state $E_{\rm 1}^{\rm rel}$.  In
\textit{model A}, $\Delta$ and $E_{\rm 1}^{\rm rel}$ are given in the
harmonic approximation by Eqs.\ \eref{eq:Delta_ho} and
\eref{eq:1st_trap_ho}, in \textit{model B} by Eqs.\ \eref{eq:Delta}
and \eref{eq:1st_trap} within a perturbative correction, respectively,
and in \textit{model C} the c.m.\ energies are calculated numerically
exact.  For given $\Delta$ and $E_{\rm 1}^{\rm rel}$ within one model,
the inelastic CIR position for a c.m.\ excitation is then determined
by \eqref{eq:crossing}.

\section{C.m.-rel.\ coupling model -- coupling strengths}
\label{sec:coupling}
After having introduced a straightforward procedure to evaluate the 
resonance positions using a model (with three different versions A, 
B, and C) for the resonance positions,
the coupling strengths are considered. They are of particular interest
for experiments, since they determine the width and thus experimental 
visibility. Furthermore, they are, e.\,g., necessary for a Landau-Zener
treatment of the dynamics at the resonances that allows for an estimate 
whether diabatic or adiabatic transitions between the involved states 
occur at the avoided level crossings. The matrix element
defining the coupling strength $W_{\mathbf{n}}$ between a bound state
$\ket{\psi^{(b)}(\mathbf{r})\, \Phi_\mathbf{n}(\mathbf{R})}$ with
c.m.\ excitation $\Delta_{\mathbf{n}}$ and the lowest trap state
$\ket{\psi_1(\mathbf{r})\, \Phi_{(0,0,0)}(\mathbf{R})}$ is
\begin{align}
  W_\mathbf{n}= \braket{\psi^{(b)}(\mathbf{r})\,
    \Phi_\mathbf{n}(\mathbf{R}) | W(\mathbf{r},\mathbf{R}) |
    \psi_1(\mathbf{r})\, \Phi_{(0,0,0)}(\mathbf{R})} \quad .
\label{eq:matrix_element}
\end{align}

For the model the wavefunctions of a harmonic confinement are adopted.
Hence, the c.m.\ wavefunction is the product of 1D harmonic oscillator
wavefunctions (here written in dimensionless units of energies in
$\hbar \omega_j$ and lengths in $\sqrt{\frac{\hbar}{M \omega_j}}$) 
\begin{align}
   \Phi_n(R_j) = \pi^{-\frac{1}{4}} \sqrt{\frac{1}{2^n n!}}\,
 \mathrm{e}^{-\frac{1}{2}R_j^2}\, \mathrm{H}_n(R_j)
 \label{eq:ho_1d}
\end{align}
where $\mathrm{H_n}(R_j)$ denote the Hermite polynomials. The c.m.\
integral in \eqref{eq:matrix_element} reduces to a sum of 1D integrals
that can be calculated even analytically.

The 3D integral over the relative-motion coordinate is more
laborious. While an expression for the trap wavefunctions for an
arbitrarily anisotropic harmonic confinement is so far (to the authors' 
knowledge) not yet known, a general solution for the trap state in a
harmonic potential with a single (but arbitrary) anisotropy,
e.\,g. $\omega_x = \omega_y =: \omega_{\perp} \ne \omega_z$ is given
in \cite{cold:idzi06}. However, the numerical evaluation of
\eqref{eq:matrix_element} with the most general version of the
relative motion wavefunction given in \cite{cold:idzi06},
\begin{align}
  \psi_{\epsilon}&(\rho,z)  =  \frac{\eta}{2 \pi^{3/2} 2^{\epsilon/2}}\, e^{-\eta \rho^2/2}\, \nonumber \\
  & \times \sum_{m=0}^{\infty} 2^{m \eta}\, \mathrm{L}_m(\eta \rho^2)\, 
  \Gamma\left(\frac{2 m \eta - \epsilon}{2}\right)\,
  \mathrm{D}_{\epsilon - 2 m \eta}\left(\sqrt{2} |z|\right),
\end{align}
has turned out prohibitively demanding. In the regime of a strongly
elongated (quasi-1D) potential the expression greatly simplifies
\cite{cold:idzi06} to
\begin{align}
  \psi_1(\rho,z) = \frac{\eta}{2 \pi^{3/2} 2^{\epsilon/2}}\, e^{-\eta
    \rho^2/2}\, \Gamma\left(-\frac{\epsilon}{2}\right)\,
  \mathrm{D}_{\epsilon}\left(\sqrt{2} |z|\right) .
\label{eq:psi1_q1d}
\end{align}
In these equations $\mathrm{D}_{\nu}$ denotes the parabolic cylinder
function, $\mathrm{L}_m$ the Laguerre polynomials, $\Gamma$ the gamma
function, and $\rho^2 = x^2 + y^2$. The wavefunctions are written in
dimensionless units of energies in $\hbar \omega_z$ and lengths in
$d_z = \sqrt{\frac{2 \hbar}{m \omega_z}}$. Moreover, the previously 
introduced definitions 
$\eta=\frac{\omega_{\perp}}{\omega_z}$ and $\epsilon =
E^{\mathrm{rel}}-\frac{\hbar}{2}(\omega_x+\omega_y+\omega_z)$ remain 
valid. Within the model (see above), the energy at the
resonance is $\epsilon = \hbar \omega_z$ (assuming that the elongation
of the trap is along the $z$ direction). It is important that for the
wavefunction the energy is \textit{not} corrected by the anharmonic
terms, because this might result in a different energy branch of the
spectrum. With $\epsilon = \hbar \omega_z$ at the resonance,
\eqref{eq:psi1_q1d} can be further simplified (in physical units) to
\begin{align}
  \psi(\rho,z)=\frac{ \sqrt{2}
    \Gamma\left(-\frac{1}{2}\right)}{4\pi^{\frac{3}{2}}d_{\perp}^2
    \sqrt d_z }\, |z|\, \exp\left(-\frac {\rho^2} {2 d_{\perp}^2}
    -\frac{z^2}{2 d_z^2}\right) 
  \label{eq:q1dwf}
\end{align}
with $d_{\perp} = \sqrt{\frac{2
    \hbar}{m \omega_{\perp}}}$ and using that for integer $n$ the relation 
$\mathrm{D}_n(x)=2^{-\frac{n}{2}}\mathrm{e}^{-\frac{x}{4}}
H_n\left(\frac{x}{\sqrt{2}}\right)$ holds.

In quasi 2D, the wavefunction can be simplified to \cite{cold:idzi06}
\begin{align}
  \psi_1(\rho,z) = \frac{1}{2 \pi^{\frac{3}{2}}}
  \mathrm{e}^{-\frac{\eta \rho^2 +z^2}{2}}
  \Gamma\left(-\frac{\epsilon}{2 \eta}\right) U\left(
    -\frac{\epsilon}{2 \eta},1,\eta \rho^2 \right)
  \label{eq:q2dwf}
\end{align}
where $U$ denotes the confluent hypergeometric function. Again,
dimensionless units of energy in $\hbar \omega_z$ and lengths in $d_z$
are used.

The rel.\ motion bound-state wavefunction \cite{cold:idzi06}
\begin{align}
  \psi_b(\mathbf{r})&= {\frac {\sqrt {{d_z}}}{{{d_{\perp}}}^{2} {(2\pi) }^{3/2}}} \nonumber \\
  & \times \int_0^{\infty}dt\, \frac{ {\rm exp}\bigg({t{\frac{ E}{\hbar\omega_z}}-{\frac
        {{z}^{2}}{2 t{{d_z}}^{2}}}- \frac{{\rho}^{2}}{2d_{\perp}^2}\coth \left( {\frac {t{{
                d_z}}^{2}}{{{ d_{\perp}}}^{2}}} \right) }\bigg) }
  {\sqrt {t} \sinh \left( {\frac {t{{ d_z}}^{2}}{{{ d_{\perp}}}^{2}}} \right)}.
\label{eq:bs_wf}
\end{align}
which is written here in in physical units, is valid for an isotropic
confinement as well as for a strongly anisotropic trap geometry. Note, 
if the energy of the bound state crosses the
energy of the trap state for positive values of the $s$-wave scattering length
that are sufficiently small ($a \lesssim d_{\perp}$), its energy at
the crossing is sufficiently small such that the bound state
wavefunction can be described in good approximation by its trap-free
counterpart
\begin{align} 
  \psi_{\rm free}(r)= \frac{1}{\sqrt{2 \pi a}}
  \frac{\mathrm{e}^{-\frac{r}{a}}}{r} \quad .
\end{align}
However, for crossings at negative $s$-wave scattering lengths, this
approximation certainly fails, because in the free case the bound
state only exists for $a > 0$. Hence, in the following
\eqref{eq:bs_wf} is used for the bound-state wavefunction.

The wavefunctions of a harmonic confinement of \textit{simple}
anisotropy are used for the model. Therefore, the rel.\ motion
integral is reduced to a two-dimensional one where the coupling term
$W(\mathbf{r},\mathbf{R})$ is averaged over the transversal direction,
$\omega_{\perp}= \frac{1}{2}(\omega_x +\omega_y)$. The matrix element
becomes
\begin{align}
  W_{\mathbf{n}} = 2 \pi \int_0^{\infty} d\rho\ \rho \int_{-\infty}^{\infty} dz\, \psi_b(\rho,z) \psi_1(\rho,z) \tilde{W}(\rho,z) 
\label{eq:matrix_element_long}
\end{align}
with 
\begin{align}
  \tilde{W}(\rho,z) = & \sum_{k=x,y} V_k  \int_{-\infty}^{\infty}dR_k\, W_k(\rho,R_k) \Phi_{n_k} \Phi_{n_0} \nonumber \\
  & + V_z \int_{-\infty}^{\infty}dZ\, W_z(z,Z) \Phi_{n_z} \Phi_{n_0} 
\end{align}
where $W(\mathbf{r},\mathbf{R}) = \sum_{j=x,y,z} V_j\,
W_j(r_j,R_j)$ and $\mathbf{n} = (n_x,n_y,n_z)$. 

Hence, different to the resonance position where three models were
introduced that differed in the treatment of the c.m.\ energies, two
models for the coupling strength were introduced that differ in the
way the rel.-motion wavefunction $\psi_1$ is treated. In quasi 1D,
\eqref{eq:q1dwf} is used for $\psi_1$ within \textit{model 1}, in
quasi 2D \eqref{eq:q2dwf} is adopted for $\psi_1$ within \textit{model
  2}. The models solve the coupling matrix element of
\eqref{eq:matrix_element} for the exact coupling term of the sextic
potential of \eqref{eq:sextic_w} with wavefunctions in the harmonic
approximation.

\section{3D-1D transition}
\label{sec:transition_1d}
In order to characterize the behavior of the resonances and to
validate the models, full six-dimensional \textit{ab initio}
calculations are performed. The numerical algorithm to solve the
stationary Schr\"odinger equation for the Hamiltonian in
\eqref{eq:Hamil} is described in \cite{cold:gris11}. For the efficient
computational treatment of ultracold atoms in optical lattices the
basis functions are symmetry adapted to the eight irreducible
representations $(A_{\rm
  g},B_{1\mathrm{g}},B_{2\mathrm{g}},B_{3\mathrm{g}},A_{\mathrm{u}},B_{1\mathrm{u}},B_{2\mathrm{u}},B_{3\mathrm{u}})$
of the orthorhombic point group $D_{\rm{2h}}$. This leads to a
corresponding block structure of the Hamiltonian matrix. As a
realistic example for an interatomic interaction potential
$U_{\mathrm{int}}$ the (numerically given) Born-Oppenheimer potential
curve of two $^7$Li atoms in the electronic state $a\, ^3\Sigma_u^+$
are used in this work, see \cite{cold:gris07} for details on how this
potential was obtained. The variation of the $s$-wave scattering
length is achieved by slight modifications of the inner-wall of the
potential \cite{cold:gris10} which effectively changes the asymptotic
behavior of the radial wavefunction and hence the $s$-wave scattering
length. Resonance positions and coupling strengths are extracted from
the \textit{ab initio} data by fitting a two-channel model to the
corresponding avoided energy crossing (see \secref{sec:2channel} in
the Appendix).

In the following, the behavior of the coupling resonances is
investigated in the transition from a 3D to a quasi-1D confinement.
Two $^7$Li atoms are considered in a sextic potential
(Eqs.\,\eref{eq:sextic_rel}-\eref{eq:sextic_w}) with
$\lambda$=1000\,nm, $\omega_x/\omega_y =1.1$, $V_y=35.9\,E_r$. To
obtain an elongation in the longitudinal $z$ direction, the potential
depth $V_z$ is decreased. Hence, an almost spherical potential is
deformed into an elongated, cigar-shaped one.

The coupling term $W$ in \eqref{eq:sextic_w} is totally symmetric and hence
only states of equal symmetry couple. Since in the following only
crossings with the first trap state without a c.m.\ excitation are
considered, only even excitations, i.\,e.\ states with even $n_j$, can
couple. Moreover, in the \textit{ab initio} calculation it suffices to
consider the $\rm A_g$ symmetry because the states involved in the
considered inelastic CIR, the first trap state, the last bound
state and the even c.m.\ excitations possess $\rm A_g$ symmetry, see
\cite{cold:gris11} for details. 

\begin{figure}[ht]
  \begin{centering}
  \includegraphics[width=0.44\textwidth]{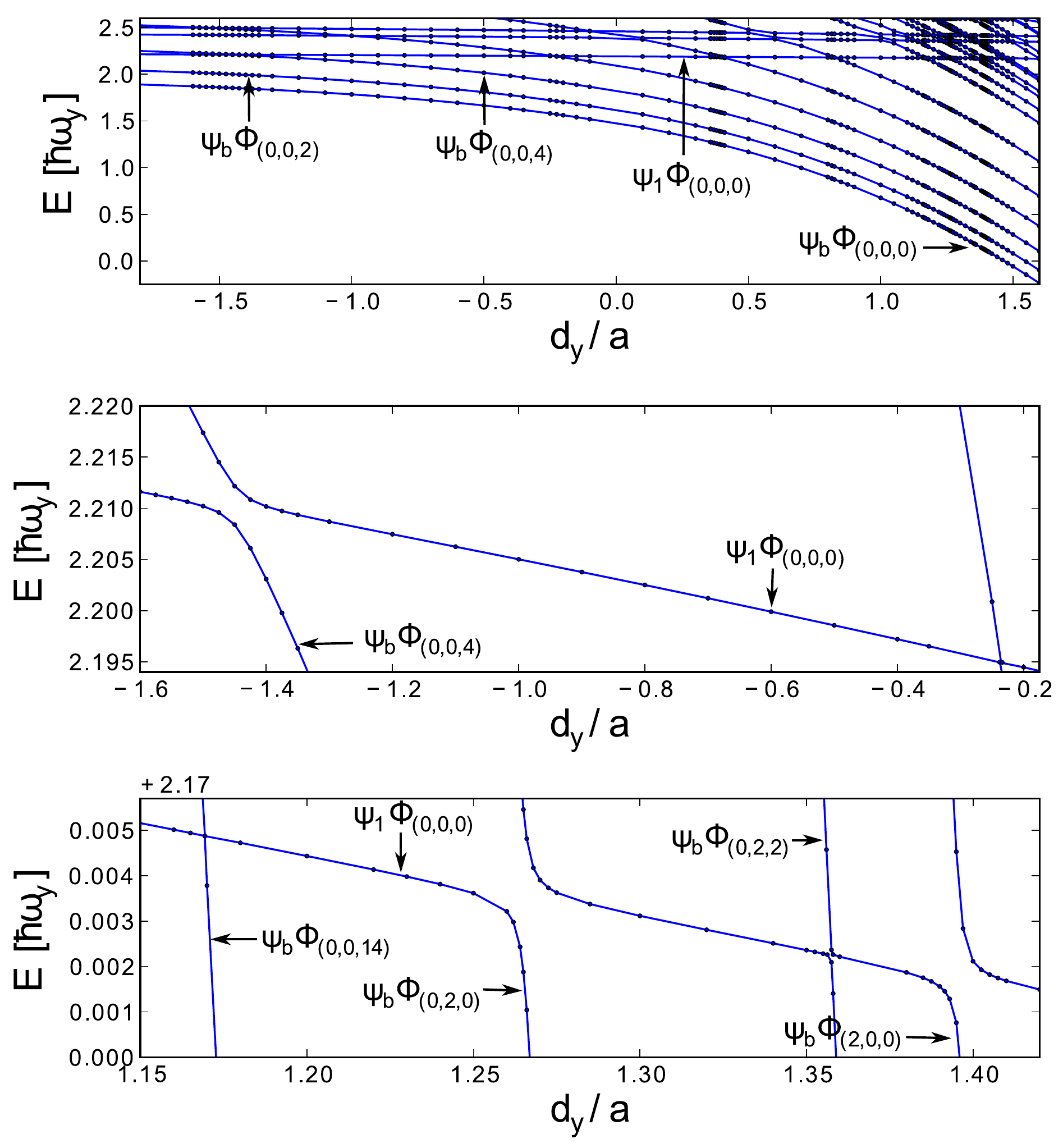}
  \caption{(Color online) \textit{Ab initio} energy spectrum of a
    $^7$Li-$^7$Li system confined in a sextic trapping potential with
    $\lambda$=1000\,nm, $\omega_x/\omega_y =1.1$, $\omega_y/\omega_z
    =10$, $V_y=35.9\,E_r$. The labels of the energy branches denote
    the corresponding diabatic states.}
  \label{fig:spec}
  \end{centering}
\end{figure}

The two lowest-order transversal resonances with $(2,0,0)$ and
$(0,2,0)$ c.m.\ excitation and the first-order longitudinal resonance
with $(0,0,4)$ c.m.\ excitation are investigated in the
following. These resonances are selected because they are the most
pronounced ones. As a rule of thumb, the coupling between a c.m.\
excited bound state and the ground trap state decreases with the order
of the c.m.\ excitation, i.\,e.\ the lowest resonances $n_i=2, n_{j
  \ne i}=0$ show the strongest coupling. A simple argument for this
rule is that the stronger oscillations in higher excited c.m.\ bound
states decrease the overlap to the trap state and hence also the
coupling matrix element \eqref{eq:matrix_element}.
\begin{table}[ht]
  \begin{center}
    \begin{tabular}{ |c|c|c|c|c|c| }
     \hline
      $\mathbf{n}$ & $(0,2,0)$ & $(0,2,2)$ & $(0,4,0)$ & $(0,6,0)$ & $(0,8,0)$ \\ 
      \hline
      $W_{\mathbf{n}}[10^{-3}\hbar \omega_y]$ &3.19 & 1.46 & 0.795  & 0.50 & 0.31 \\
      \hline
    \end{tabular}
  \end{center}
  \caption{Coupling strengths $W_{\mathbf{n}}$ for $\omega_x/\omega_y = 1.1$,  $\omega_y/\omega_z = 1$, $V_y=35.9\,E_r$ of the first trap state with c.m.\ excited bound states for different c.m.\ excitations $\mathbf{n}$ obtained with the model 1. }
  \label{tab:tab_one}
\end{table}
Numerically, this rule is verified by calculating the coupling matrix
element \eqref{eq:matrix_element} for different c.m.\ excitations, see
\tabref{tab:tab_one}, confirming that $W_{\mathbf{n}}$ decreases with
an increasing order of the c.m.\ excitation. It can also be seen in
the energy spectrum in \figref{fig:spec} where high-order resonances
do not show visible avoided crossings \footnote{The reason why the
  $(0,0,4)$ resonance is much stronger then the, e.\,g., $(0,2,2)$
  resonance in \figref{fig:spec} compared to the \tabref{tab:tab_one}
  is the different confinement geometry, strongly elongated against
  almost isotropic, respectively.}. The fact that the coupling
decreases with the c.m.\ excitation of the bound state delivers an
explanation why in \cite{cold:bold05} it is (correctly) concluded that
the dimer-production rate at an c.m.-rel.\ coupling resonance
involving a very highly c.m.-excited bound state is negligible.

While lowest-order resonances show the strongest coupling, in the
following the lowest-order longitudinal resonance with c.m.\
excitation $(0,0,2)$ is not considered because its position fades away
to $\frac{d_y}{a} \ll -1$ with decreasing $\omega_z$ where the bound
state is very shallow and has lost its characteristic small
interatomic distance. In the full spectrum the crossing can therefore
not be easily resolved any more. This can be seen in \figref{fig:spec}
in the upper most panel, where the energy of the $\ket{\psi_b
  \Phi_{(0,0,2)}}$ state asymptotically approaches $\ket{\psi_1
  \Phi_{(0,0,0)}}$ without a pronounced crossing.

\begin{figure}[ht]
  \begin{centering}
    \includegraphics[width=0.44\textwidth]{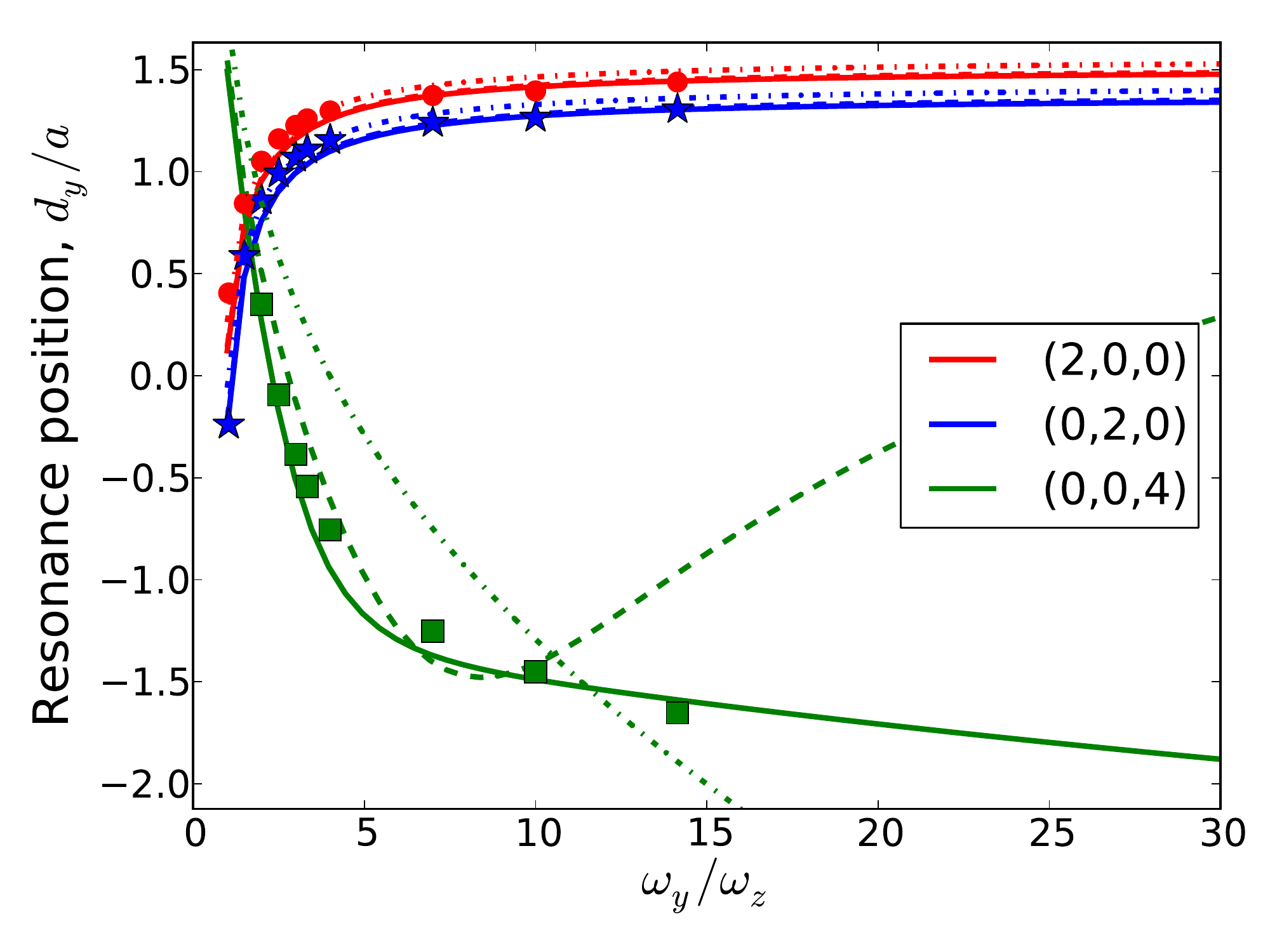}
    \caption{(Color online) Resonance positions for different c.m.\
      excitations $\mathbf{n} = (n_x,n_y,n_z)$ (see legend) obtained
      by full \textit{ab initio} calculations [dots (2,0,0), stars
      (0,2,0), squares (0,0,4)] and the model
      (lines) if the tightness in $z$ direction is reduced. The dashed-dotted
      lines correspond to the harmonic approximation (model
      A). The dashed lines are obtained by a perturbative correction of
      the energies (model B). For the solid lines, the correction
      is calculated in a numerically exact way (model C).}
  \label{fig:1d_eta10}
  \end{centering}
\end{figure}

In \figref{fig:1d_eta10} the three considered resonance positions of
the models with different degrees of corrections together with
\textit{ab initio} results are shown. Before discussing the validity
of the models, the behavior of the resonances is analyzed. For small
anisotropies $\omega_y/\omega_z \lesssim 2$ the $(0,0,4)$ resonance
lies at larger values of $d_y/a$ (the green curve lies above the red
and blue curves in \figref{fig:1d_eta10}) simply because the c.m.\
excitation $\Delta_{(0,0,4)}$ is larger than $\Delta_{(0,2,0)}$ and
$\Delta_{(2,0,0)}$. Therefore, the c.m.\ excited bound state
$\ket{\psi_b \Phi_{(0,0,4)}}$ crosses at larger values of $d_y/a$ than
the ones with a single excitation. A decrease of $V_z \propto
\omega_z^2$ decreases the spacings of the states that have a c.m.\
excitation in the $z$ direction. Hence, for decreasing $\omega_z$ the
$(0,0,4)$ resonance crosses constantly at smaller values of $d_y/a$
which explains the monotonic decrease of the \textit{ab initio}
results in \figref{fig:1d_eta10} (green squares). On the other hand,
the transversal c.m.\ excitations ($\Delta_{(0,2,0)}$ and
$\Delta_{(2,0,0)}$) remain constant for a variation of $\omega_z$. Yet, a
decrease in $\omega_z$ also decreases the energy of the first trap
state. Therefore, the transversally excited c.m.\ states $\ket{\psi_b
  \Phi_{(2,0,0)}}$ and $\ket{\psi_b \Phi_{(0,2,0)}}$ cross at larger
values of $d_y/a$ with decreasing $\omega_z$ converging to a finite
value as $\omega_z \to 0$.

Next, the validity of the models is considered. For the transversal
$(2,0,0)$ and $(0,2,0)$ resonances the perturbative corrections Eqs.\
\eref{eq:Delta} and \eref{eq:1st_trap} agree with the numerically
exact corrections (the dashed and solid lines are
indistinguishable). The resonance positions resulting from the
harmonic approximation (dashed-dotted lines, model A) show an almost
constant offset towards larger values compared to model B were the
anharmonicity in the c.m.\ motion has been taken into account. This
small offset is due to the missing negative quartic term that is
present for the sextic potential. Certainly, the models give very good
quantitative agreement with the \textit{ab initio} calculations. For
strong anisotropies the results of the models B and C (dashed and
solid lines) are in perfect quantitative agreement, e.\,g., at
$\omega_y/\omega_z = 10$ the models give $d_y/a = 1.397 $ and the
\textit{ab initio} method results in $d_y/a = 1.396$ for the $(2,0,0)$
resonance. This excellent applicability of the model for the resonance
positions in the quasi-1D regime lead to the quantitative agreement of
the model compared to the loss resonances measured in
\cite{cold:hall10b} as shown in \cite{cold:sala12}.

The results for the longitudinal $(0,0,4)$ resonances are more
sensitive. First, with a decreasing potential depth the anharmonicity
is important already for low lying states. Second, for the $(0,0,4)$
resonance higher c.m.\ excitations are involved which enhances the
influence of the anharmonicity.  Moreover, for a decreasing resonance
position the bound state crosses the trap state with an increasingly
comparable slope which makes the position more sensitive to energy
variations. Therefore, model A including uncorrected, harmonic c.m.\
excitation is inaccurate over the entire range of anisotropies.  The
perturbatively corrected model B is acceptable for mild anisotropies
($\omega_y/\omega_z \lesssim 10$) but has a wrong behavior for
$\omega_y/\omega_z \gtrsim 10$. Finally, model C which corrects the
c.m.\ excitations and the energy of the trap state exactly numerically
is quantitatively accurate over the entire range of the scattering
length, even in the limit $\omega_z \to 0$ (green solid line).

The quantitative accuracy for very large values of the anisotropy,
$\omega_y/\omega_z \gtrsim 10$, could not be confirmed by the \textit{ab
  initio} method \cite{cold:gris11}. The basis set of the method
consists of spherical harmonics which are not well suited for
resolving extremely anisotropic wavefunctions unless high
angular-momentum quantum numbers are employed which leads to an
inconvenient scaling of the numerical effort.

\begin{figure}[ht]
  \begin{centering}
    \includegraphics[width=0.44\textwidth]{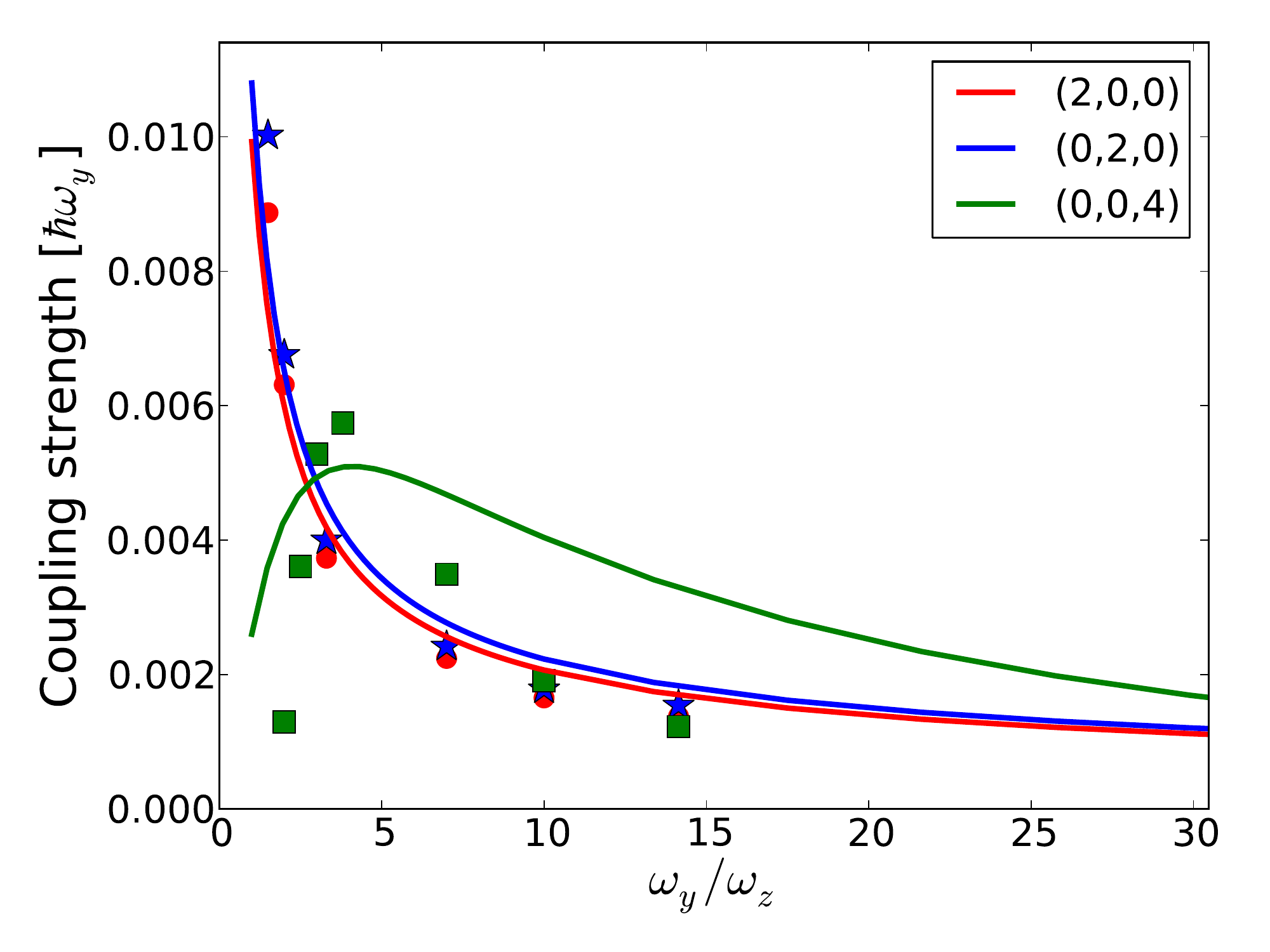}
    \caption{(Color online) Coupling strength for different c.m.\
      excitations $\mathbf{n} = (n_x,n_y,n_z)$ (see legend) obtained
      by full \textit{ab initio} calculations [dots (2,0,0), stars
      (0,2,0), squares (0,0,4)] and the model 1 (solid lines).}
  \label{fig:1d_eta100_coup}
  \end{centering}
\end{figure}

In \figref{fig:1d_eta100_coup} the coupling strengths corresponding to
\figref{fig:1d_eta10} are shown. Again, the overall behavior is
discussed prior to the validity of the model 1.  The coupling strength
of the transversal resonances with $(2,0,0)$ and $(0,2,0)$ c.m.\
excitations (red circles and blue stars, respectively) decreases
monotonically with increasing anisotropy but approaches a constant
value for $\omega_z \to \infty$. Otherwise the observation of the
$(2,0,0)$ and $(0,2,0)$ resonances in \cite{cold:hall10b} where the
anisotropy of the external confinement is $\approx 10^3$ would not
have been possible. A simple argument for the monotonic decrease is
that $W(\mathbf{r},\mathbf{R}) \propto V_k$.

For the resonance with longitudinal $(0,4,0)$ c.m.\ excitation, a
non-monotonic behavior is visible. In the limit of $\omega_z \to 0$
which corresponds to a zero potential in the $z$ direction, the
coupling of the resonances with a c.m.\ excitation in the longitudinal
direction vanishes. This is intuitive, since without a confinement
potential there exists no confinement-induced c.m.-rel.\ coupling. On
the other hand, a decrease in the potential depth $V_z$ leads to an
enhancement of the anharmonicity-induced coupling, since the potential
becomes more anharmonic (this will be discussed in more detail in
\secref{sec:opt}). The result of these counter-acting effects is the
non-monotonic curve with the local maximum for the $(0,0,4)$ resonance
and a vanishing coupling for $\omega_z \to 0$.

Next, the validity of model 1 is considered. For the longitudinal
resonance, model 1 provides the correct qualitative behavior and
reproduces the local maximum accurate in position. In general,
however, it does not provide highly accurate quantitative
agreement. Especially for larger anisotropies ($\omega_y/\omega_z
\gtrsim 7$), the model 1 overestimates the coupling strengths. Again,
this behavior is understandable since for the higher-order
longitudinal resonances the anharmonicity becomes increasingly
important, which cannot be modeled accurately with wavefunctions in
the harmonic approximation.

For the transversal resonances with $\Delta_{(0,2,0)}$ and
$\Delta_{(2,0,0)}$ c.m.\ excitation, the coupling strengths predicted
by the model 1 agree quantitatively very well with the \textit{ab
  initio} ones for $\omega_y/\omega_z \ge 2$. This agreement is
remarkable, since no free parameters were used in the models.

\section{3D-2D transition}
\label{sec:transition_2d}

In the following, the transition from a 3D to a quasi-2D confinement
is considered. Again, as a realistic example two $^7$Li atoms in 
a sextic confinement of
$\lambda$=1000 nm and $V_y$=35.9$E_r$ are chosen.  To obtain a 
pancake-shaped trap $\omega_x$ and $\omega_z$ are decreased, keeping the ratio
$\omega_x/\omega_z = 1.1$ constant.
The lowest-order resonance with $(0,2,0)$ c.m.\ excitation and the
next to leading order resonances with $\mathbf{n}=(4,0,0)$ and
$\mathbf{n}=(0,0,4)$ are considered. Again, in analogy to the 1D case
these resonances are the most pronounced ones, having in mind that the
two lowest-order resonances with excitations in the weakly confined
directions, i.\,e.\ with c.m.\ excitations $(2,0,0)$ and $(0,0,2)$,
fade away towards large negative values of $d_y/a$, loosing their
resonant character.

\begin{figure}[ht]
  \begin{centering}
    \includegraphics[width=0.44\textwidth]{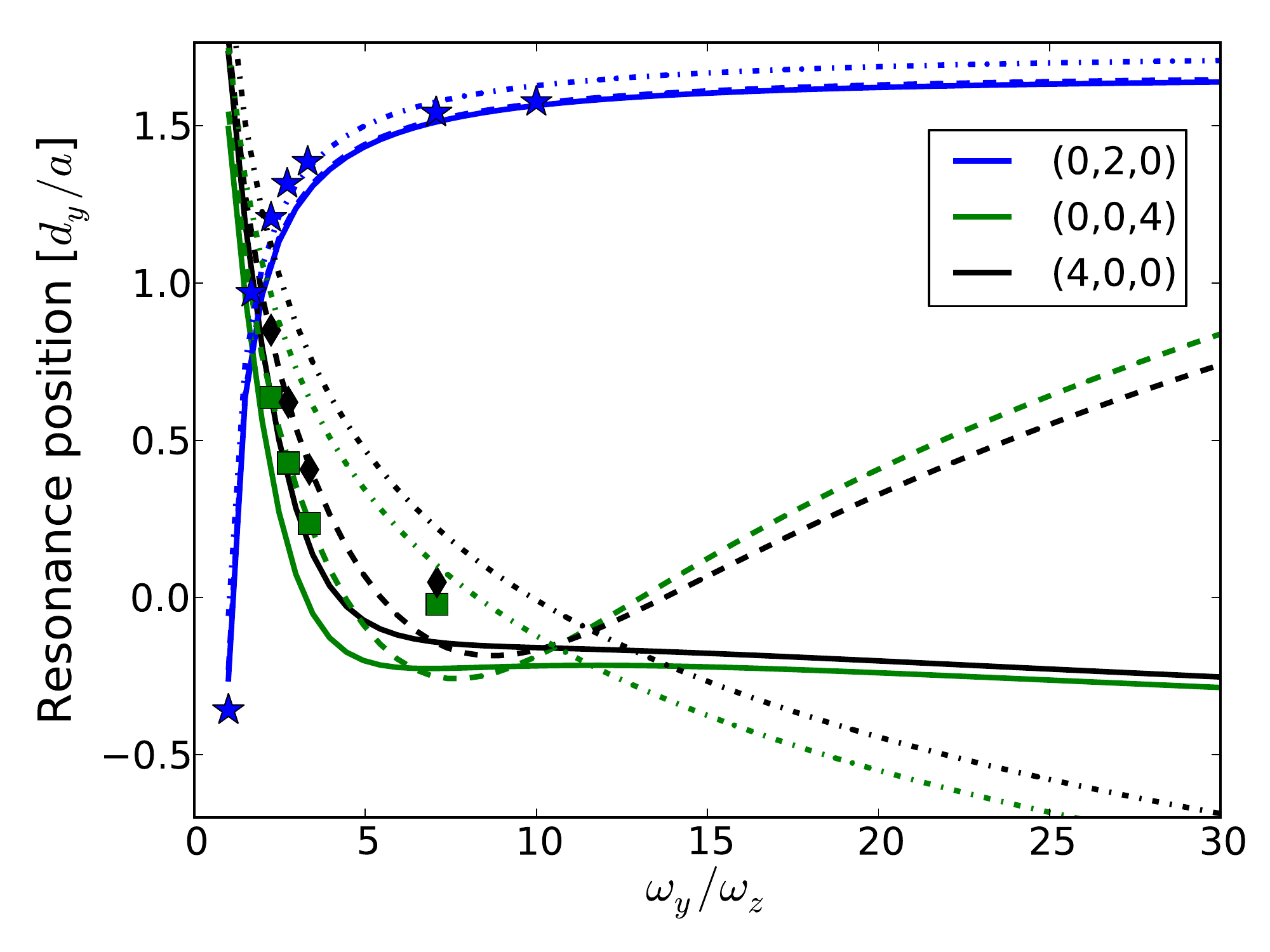}
    \caption{(Color online) Resonance positions for different c.m.\
      excitations obtained by full \textit{ab initio} calculations
      (dots) and the model (lines). The dashed lines are obtained by
      the perturbative correction of the energies (model B), the solid
      lines indicate the numerically exact c.m.\ correction (model C)
      and the dashed-dotted lines correspond the harmonic
      approximation (model A).}
  \label{fig:2d_eta100}
  \end{centering}
\end{figure}

As before, the behavior of the resonance positions in
\figref{fig:2d_eta100} is discussed first based on the \textit{ab
  initio} results. A similar behavior as for the 1D case is
visible. By the same arguments that hold in the 3D to 1D transition,
the $(0,2,0)$ resonance with excitation in the strongly confined
direction starts at negative values of $d_y/a$ for small anisotropies
and converges to an asymptotic value for strong anisotropies
($\omega_x, \omega_z \to 0$). The higher order $(4,0,0)$ and $(0,0,4)$
resonances start at positive values of $d_y/a$ at small anisotropies
and do not converge to an asymptotic value for $\omega_x,\omega_z \to
0$.

For the $(0,2,0)$ resonance, the model A is again shifted to slightly higher resonance
positions due to the absence of a negative quartic term. While for
intermediate anisotropies the harmonic theory gives slightly better
quantitative agreement to the \textit{ab initio} calculations, the
asymptotic value is quantitatively reproduced to high accuracy within
the corrected model B, where a perturbative treatment of the energy
corrections is sufficient.

For the higher order $(4,0,0)$ and $(0,0,4)$ resonances, the
perturbative treatment of the corrections gives an almost perfect
quantitative description of the resonance positions for mild
anisotropies. However, for strong anisotropies it fails (shows a
minimum in the resonance positions and then goes to positive values of
$d_y/a$) and the exact treatment of the 1D c.m.\ excitation within
model C delivers the most accurate results.

\begin{figure}[ht]
  \begin{centering}
    \includegraphics[width=0.44\textwidth]{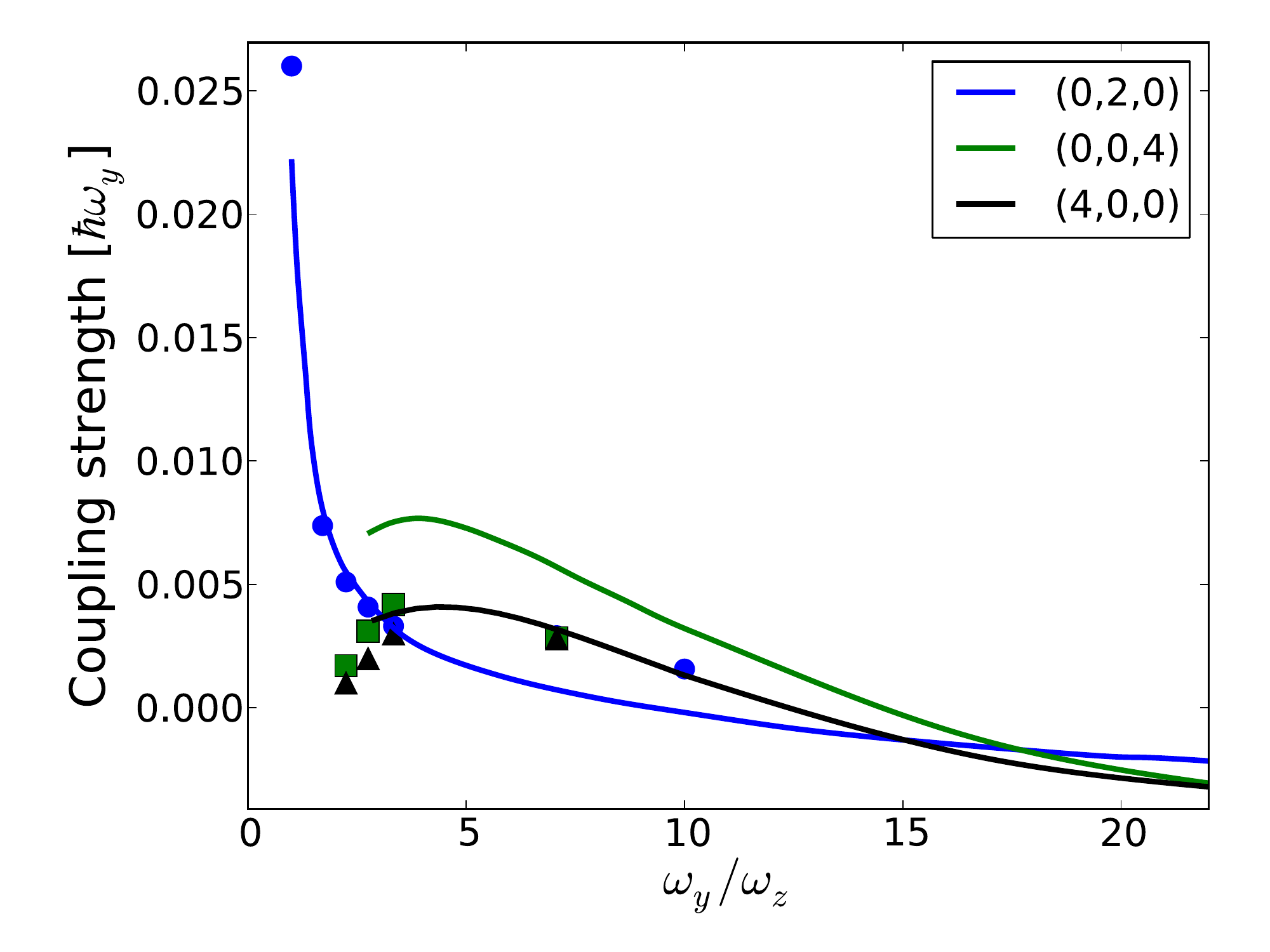}
    \caption{(Color online) Coupling strength for different c.m.\
      excitations $\mathbf{n} = (n_x,n_y,n_z)$ (see legend) obtained
      by full \textit{ab initio} calculations [dots (0,2,0), squares
      (0,0,4), triangles (4,0,0)] and the model 2 (solid lines). }
  \label{fig:coup_2d_eta100}
  \end{centering}
\end{figure}

In \figref{fig:coup_2d_eta100} the coupling strengths for the
transition from a 3D to a quasi-2D confinement are shown. The
\textit{ab initio} results show a constant decrease for the $(0,2,0)$
resonance. This is in analogy to the resonances with excitations in the
strongly confined direction in the transition to a cigar-shaped
potential shown in \figref{fig:1d_eta100_coup}. Again, the decrease
can be explained by the decrease of the coupling potential
$W(\mathbf{r},\mathbf{R}) \propto V_k$ if $V_x$ and $V_z$ are reduced
in the transition to a pancake-shaped confinement.

The behavior of the coupling strengths of the $(4,0,0)$ and $(0,0,4)$
resonances in a pancake-shaped confinement exhibits a similar behavior
as the longitudinal $(0,0,4)$ resonance in a cigar-shaped confinement
shown in \figref{fig:1d_eta100_coup}. The \textit{ab initio} results
demonstrate that the coupling strength is close to zero for an almost
isotropic confinement (that is why it cannot be resolved for
$\omega_y/\omega_z \lesssim 2$), increases until it reaches a maximum
and then falls off to zero as $\omega_x,\omega_z \to 0$. Again, its
behavior is a result of the counter-acting effect that on one hand  
decreasing the potential depth increases the anharmonicity and hence
the coupling strength, but on the other hand $\omega_z \to 0$ corresponds to
switching off the confinement leading to a vanishing confinement-induced
coupling.

A breakdown of model 2 is detected for very large anisotropies
($\omega_y/\omega_z > 10$) where the model 2 predicts negative
coupling strengths for all resonances.  The reason why the coupling
integral in \eqref{eq:matrix_element_long} can result in negative
values is the negative quartic term in \eqref{eq:sextic_w}. However,
while negative coupling strengths for themselves are not a problem
yet, an unphysical discontinuity is introduced when taking the absolute
value. Hence, the sign change (or even vanishing value) of the coupling
strength is unphysical. Since such a behavior is absent for
the cigar-shaped regime, the used harmonic quasi-2D trap state
wavefunction, \eqref{eq:q2dwf} turns out to be inappropriate here.

Still, the model 2 reproduces correctly the decreasing coupling
strength for the $(0,2,0)$ resonance. For smaller anisotropies
$\omega_y/\omega_z \lesssim 5$ it is even quantitatively accurate. For the
$(4,0,0)$ and $(0,0,4)$ resonances, the non-monotonic behavior is
reproduced qualitatively.

In general, for the positions as well as for the coupling strengths,
the models in quasi 1D show a better quantitative agreement than the
corresponding models in quasi 2D, simply because for a single
decreasing potential depth the anharmonicity effects are milder
compared to the pancake-shaped potential and can be reproduced by the
model that is based on the harmonic approximation more accurately.

\section{Simultaneous variation of the potential depth}
\label{sec:opt}

It is an important question how the c.m.-rel.\ coupling can be
optimized. As demonstrated above, the coupling at the lowest-order
resonances, i.\,e.\ with c.m.\ excitations $n_i=2, n_{j \ne i}=0$, 
and with excitations in the tightly confined direction have a peak
coupling strength for an isotropic trap and then monotonically
decrease with the anisotropy, i.\,e.\ with a decreasing potential
depth in the weakly confined direction(s).

Higher-order resonances in the weakly confined direction(s) have a very
small coupling for an isotropic confinement, peak at mild anisotropies
and then decrease to zero for increasing anisotropies.

\begin{table}[ht]
  \begin{center}
    \begin{tabular}{ |c|c|c|c|c|c| }
      \hline
      $V_y/E_r$ &53.9& 36.0& 18.0& 9.0& 4.5  \\ 
      \hline
      $W_{\mathrm{ab\ initio}}[10^{-3}\hbar \omega_y]$ &7.7 & 9.3 &  13.2 & 19.0 & 25.0 \\ 
      \hline
      $W_{\mathrm{model}}[10^{-3}\hbar \omega_y]$ &5.2 & 5.9 & 6.7 & 6.3 & 2.6 \\
      \hline
    \end{tabular}
  \end{center}
  \caption{Coupling strengths $W_{\mathbf{n}}$ for $\omega_x/\omega_y = 1$, $\omega_y/\omega_z = 0.5$ of the $(0,2,0)$ resonance for different values of the potential depth for \textit{ab initio} calculation and model 1. }
  \label{tab:tab2}
\end{table}

In general, the coupling strength can also be modified by a
simultaneous variation of the potential depth, i.\,e.\ by a variation
of all potential depths and not only selected ones leading to a
different (quasi-1D or quasi-2D) trap geometry. The \textit{ab initio}
results in \tabref{tab:tab2} demonstrate that the coupling increases
with a decrease of the potential depth $V$. This behavior can be
understood intuitively. As the potential becomes deeper, the harmonic
approximation becomes more accurate and it has a zero c.m.-rel.\
coupling. While the model follows this behavior for a deep potential,
it looses the accuracy, if the potential gets to shallow, and results in an
unphysical decrease of the coupling. The reason for this failure of
the model is that the harmonic wavefunctions become less accurate for 
a decreasing potential depth.

In the case of an optical lattice, the decrease of the potential depth
has yet another important consequence. A tunnel coupling between
neighboring wells enhances drastically the c.m.-rel.\ coupling. In
fact, \textit{ab initio} calculations of a double and a quadruple-well
potential have shown that in this case even high-order resonances show
a considerable coupling. Moreover, the results of the calculation
explain some of recently measured loss resonances in a shallow 3D optical
lattice in an ultracold gas of Cs atoms \cite{cold:mark15}.

\begin{table}[ht]
  \begin{center}
    \begin{tabular}{ |c|c|c|c|c|c| }
      \hline
      $V_y/E_r$ &53.9& 36.0& 18.0& 9.0& 4.5  \\ 
      \hline
      $d_y/a_{\mathrm{ab\ initio}}$ &0.9& 0.91& 0.92& 0.9& 0.81 \\ 
      \hline
      $d_y/a_{\mathrm{model}}$ & 0.82 & 0.8& 0.77& 0.72& 0.66\\
      \hline
    \end{tabular}
  \end{center}
  \caption{Resonance position for $\omega_x/\omega_y = 1$, $\omega_y/\omega_z = 2$ of the $(0,2,0)$ resonance for different values of the potential depth for \textit{ab initio} calculations and the model B with the perturbative energy correction. The harmonic approximation of the model A predicts a resonance position of $d_y/a=0.886$ independent of $V$.  }
  \label{tab:tab3}
\end{table}

With a decreasing potential depth the harmonic approximation of the
potential becomes inaccurate and the limitations of the introduced
models become visible. In \figref{fig:1d_eta10} a good agreement of the
harmonic approximation was visible for the positions, especially for
small anisotropies. In the harmonic approximation, i.\,e.\ without the
energy corrections in Eqs.\,\eref{eq:Delta} and \eref{eq:1st_trap}, the
position of the resonances is independent of the potential
depth. However, with decreasing potential depth, the anharmonic terms
in the sextic potential start to have a significant influence already
at energies of the lowest trap state and hence influence the resonance
positions. In \tabref{tab:tab3} the dependence of the position for a
mild anisotropy is compared for the model B and \textit{ab initio}
calculations. While the \textit{ab initio} results are almost constant
for a deep potential, the resonance position decreases for a decreasing
potential depth to small values. While the model A in the harmonic
approximation is independent of the potential depth, the
energy-corrected model B reflects this decrease reasonably.

\section{Wavefunction analysis}
\label{sec:wf}

C.m.-rel.\ coupling resonances have been directly observed
experimentally in a two-body system via coherent molecule formation
\cite{cold:sala13} and indirectly in a many-body system in terms of
particle loss and heating \cite{cold:hall10b,cold:sala12}. The latter
is a consequence of the molecule formation at the resonance. Both
measurements detected resonances where the bound state was excited in
a strongly confined direction, i.\,e.\ in quasi-1D the $(2,0,0)$ and
$(0,2,0)$ resonances \cite{cold:sala13,cold:hall10b}, and in quasi-2D
the $(0,2,0)$ resonance \cite{cold:hall10b} was detected. Due to the
anisotropy of the confinements, the resonance position occurred for
positive values of the $s$-wave scattering length. It was demonstrated
above, see, e.\,g.\ \figref{fig:1d_eta10}, that also resonances at
negative values of the $s$-wave scattering length occur: for
lowest-order resonances if the anisotropy is kept small, or for
higher-order resonances for a stronger anisotropy of the confinement.

\begin{figure}[ht]
  \begin{centering}
  \includegraphics[width=0.44\textwidth]{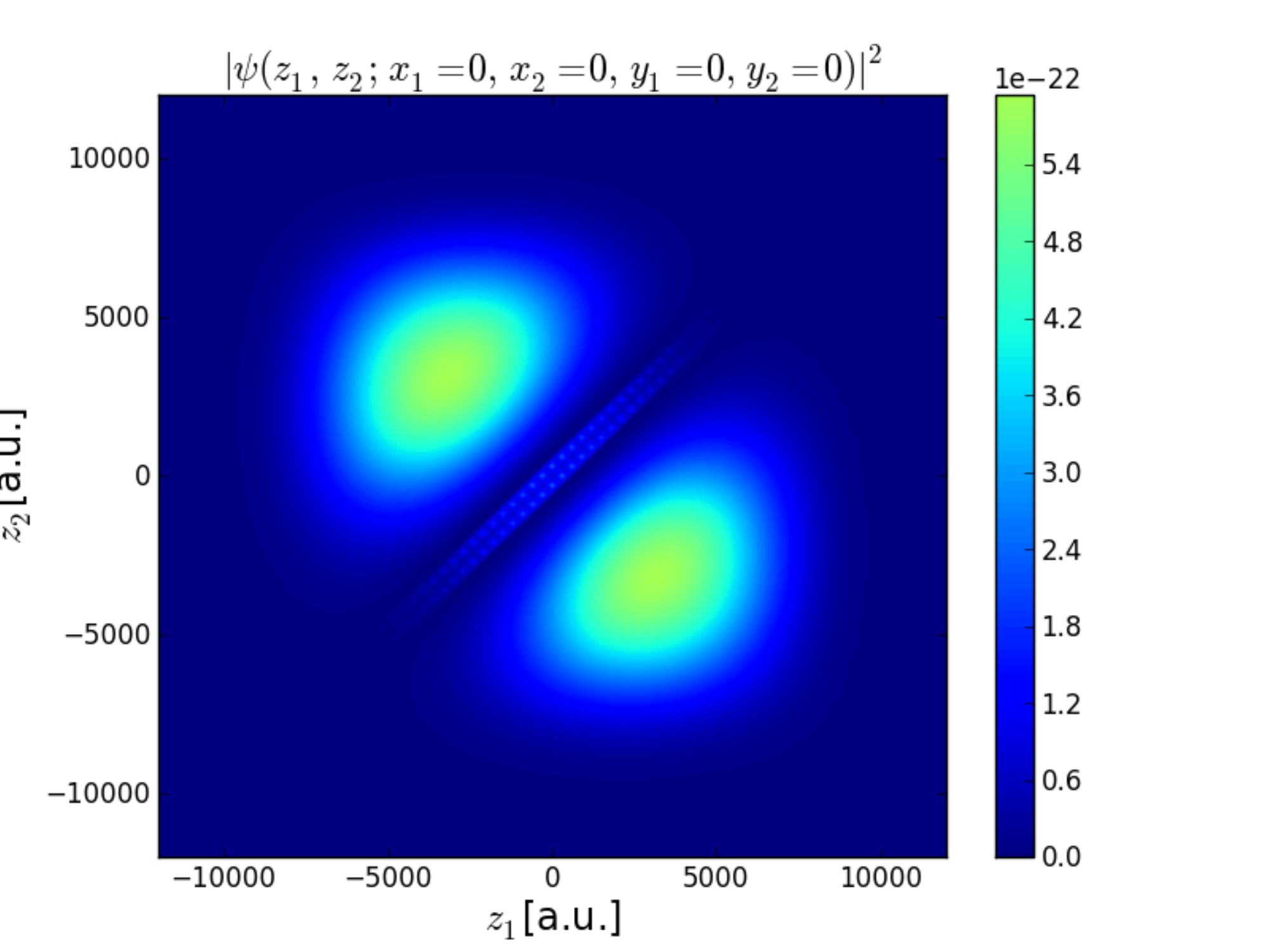}\newline
  \includegraphics[width=0.44\textwidth]{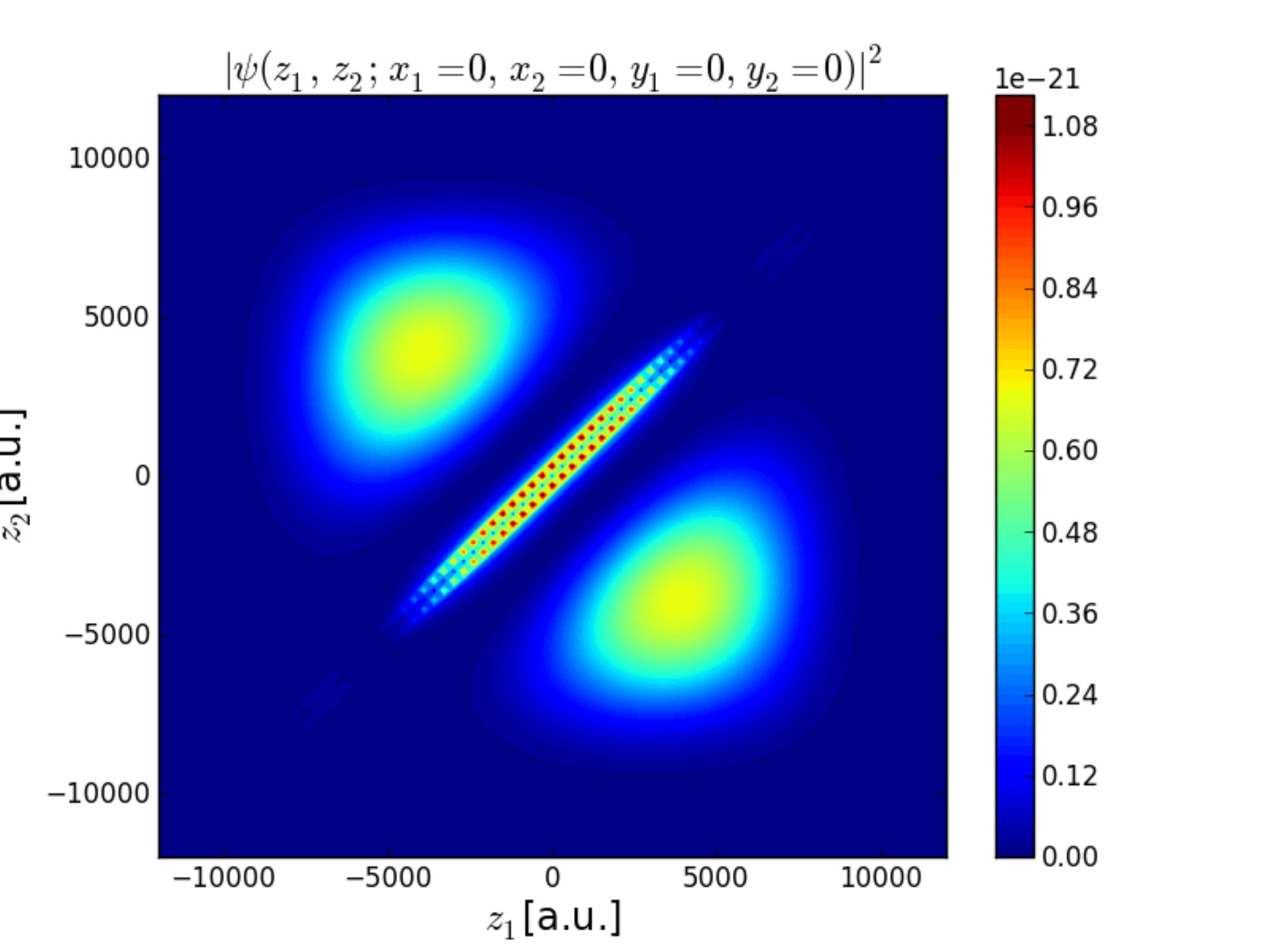}
  \caption{(Color online) Cuts along the elongated z-direction
    ($\left|\Psi(z_1,z_2;x_1=x_2=y_1=y_2=0)\right|^2$) through the full
    six-dimensional \textit{ab initio} wavefunction.\\
    Upper part: The ground trap state diabatically described by
    $\ket{\psi_1 \Phi_{(0,0,0)}}$ at $d_y/a
    = 1.42$.\\
    Lower part: The ground trap state diabatically described by
    $\ket{\psi_1 \Phi_{(0,0,0)}}$ at $d_y/a = -1.2$.\\
    (For both plots an identical color code has been used).}
  \label{fig:cntr_trap}
  \end{centering}
\end{figure}

\begin{figure}[ht]
  \begin{centering}
  \includegraphics[width=0.44\textwidth]{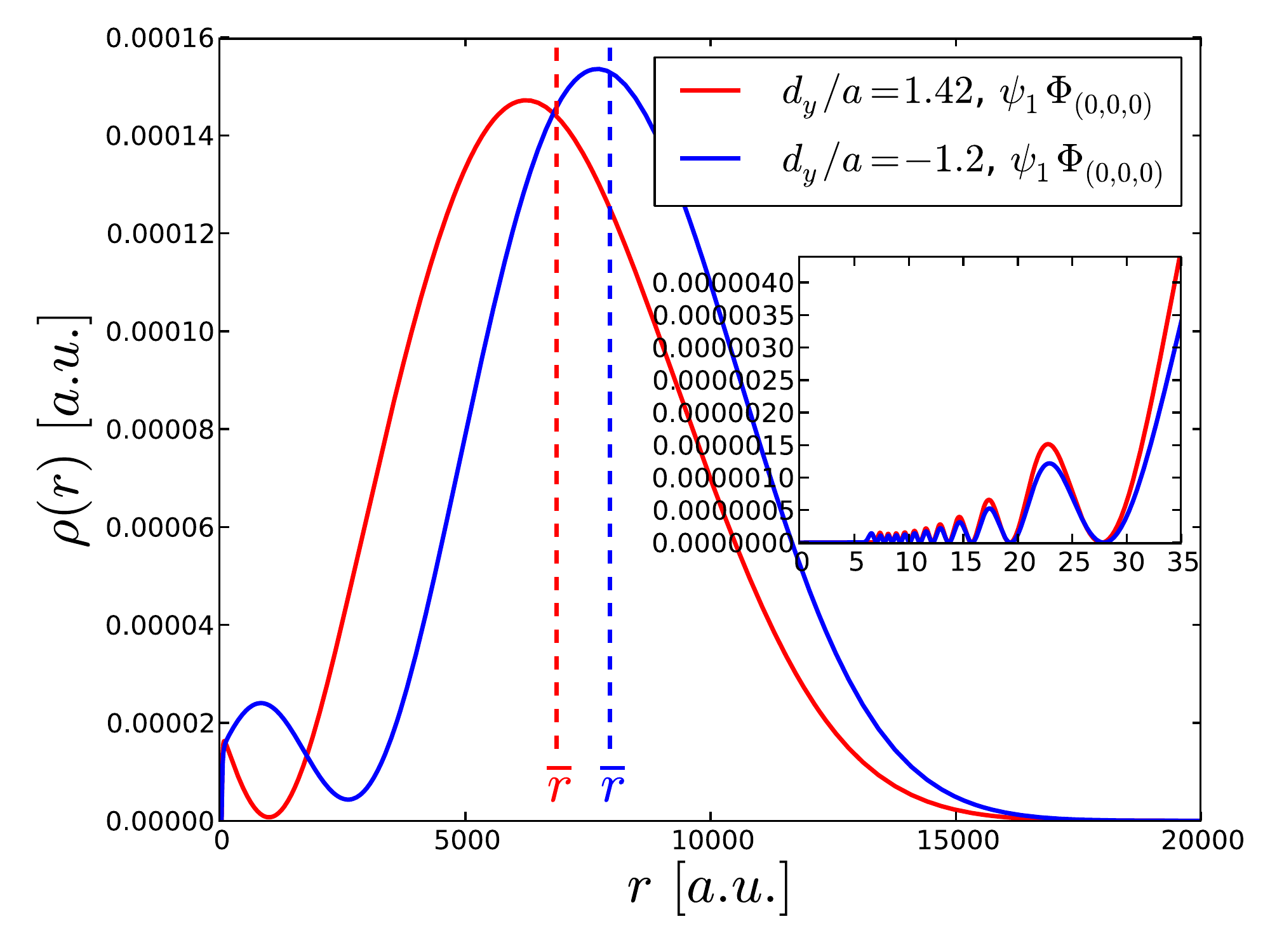}
  \caption{(Color online) Radial pair density of the first trap state
    for different values of the $s$-wave scattering length.}
  \label{fig:raddens_trap}
  \end{centering}
\end{figure}

In the following, \textit{ab initio} wavefunctions are analyzed for the
system of two $^7$Li atoms confined to a sextic potential with
parameters $\lambda$=1000\,nm, $\omega_x/\omega_y =1.1$,
$\omega_x/\omega_y =10$, $V_y=35.9\,E_r$. The corresponding energy
spectrum is shown in \figref{fig:spec}. Considered are the densities
of the bound and trap states involved in the transversally excited
$(2,0,0)$ resonance and the longitudinally excited $(0,0,4)$
resonance. The positions at which the wavefunctions are investigated are
chosen such that the overlaps of the involved trap and the bound states
are still small in order to compare the characteristics of the
states. This is $d_y/a=1.42$ for the $(2,0,0)$ and $d_y/a=-1.2$ for
the $(0,0,4)$ resonance, respectively.

In \figref{fig:cntr_trap} cuts through the trap-state densities are
shown. Since both states have the same diabatic state, i.\,e.\
$\ket{\psi_1 \Phi_{(0,0,0)}}$, they have the same global nodal
structure, i.\,e.\ two regions of large probability to the find the
particles separated from each other, away from the diagonal
$z_1=z_2$. This can also be seen considering the mean radial density
\begin{align}
  \overline{r} = \int_0^{\infty} \mathrm{d}r \, r\, \rho(r).
  \label{eq:mean1}
\end{align}
which is a measure for the mean distance of the particles. In \eqref{eq:mean1},
\begin{align}
  \rho(r) = r^2 \, \int \, \mathrm{d}V_\mathbf{R} \, \mathrm{d}\Omega_\mathbf{r} \,
  | \Psi(\mathbf{r}, \mathbf{R}) |^2
  \label{eq:raddens}
\end{align}
is the radial pair density where $\Psi(\mathbf{r}, \mathbf{R})$
denotes the full six-dimensional wavefunction of the system,
$\mathrm{d}V_\mathbf{R}$ is the c.m.\ volume element and
$\mathrm{d}\Omega_\mathbf{r}$ is the angular volume element of the
rel.\ motion. For the trap states of \figref{fig:cntr_trap} the radial
pair density is shown in \figref{fig:raddens_trap}. The large
probability for the particles to be off-diagonal in
\figref{fig:cntr_trap} are clearly reflected. This can be quantified
by the mean radial distance which is $\overline{r_{t}} = 3.95\,d_{y} =
1.25 d_z$ at $d_y/a = 1.42$ and $\overline{r_{t}} = 4.56\,d_{y} =
1.44\,d_z$ 
at $d_y/a = -1.2$. Hence, the mean distance of the trap state is on
the order of the longitudinal trap length $d_z = 291\,$nm which
reflects the elongated shape of the confinement.

In the region of interaction a strong and small-scale nodal structure
is visible close to the diagonal $z_1=z_2$ where both particles are
close to each other. The nodal structure is also visible in the inset
of the radial pair-density plot which shows $\rho(r)$ for small
$r$. In this region, the Born-Oppenheimer interaction potential
possesses a deep minimum (compared to the energy scale of the trapping
potential) which supports many bound states leading to the rich nodal
structure.

\begin{figure}[ht]
  \begin{centering}
  \includegraphics[width=0.44\textwidth]{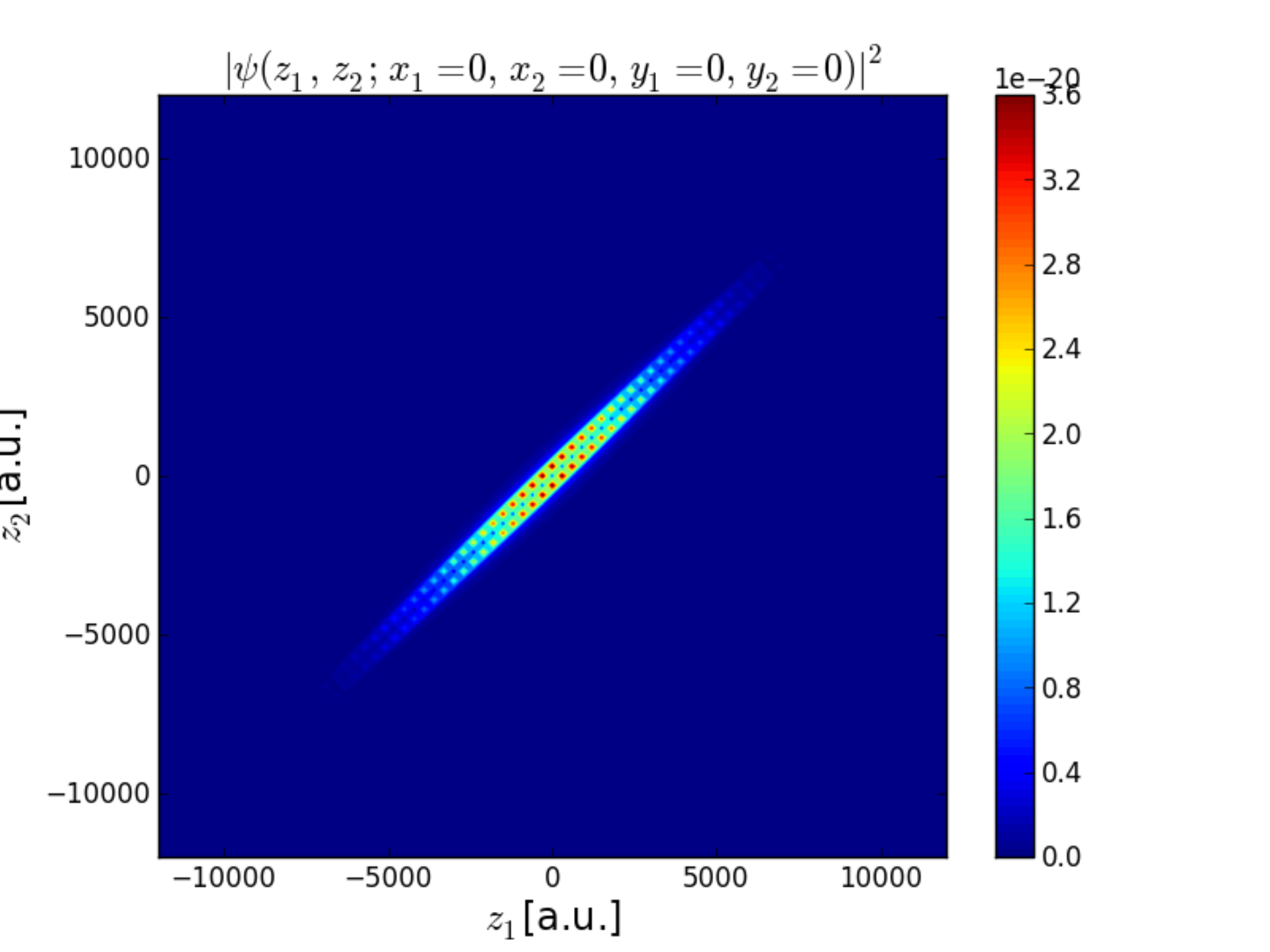}\\
  \includegraphics[width=0.44\textwidth]{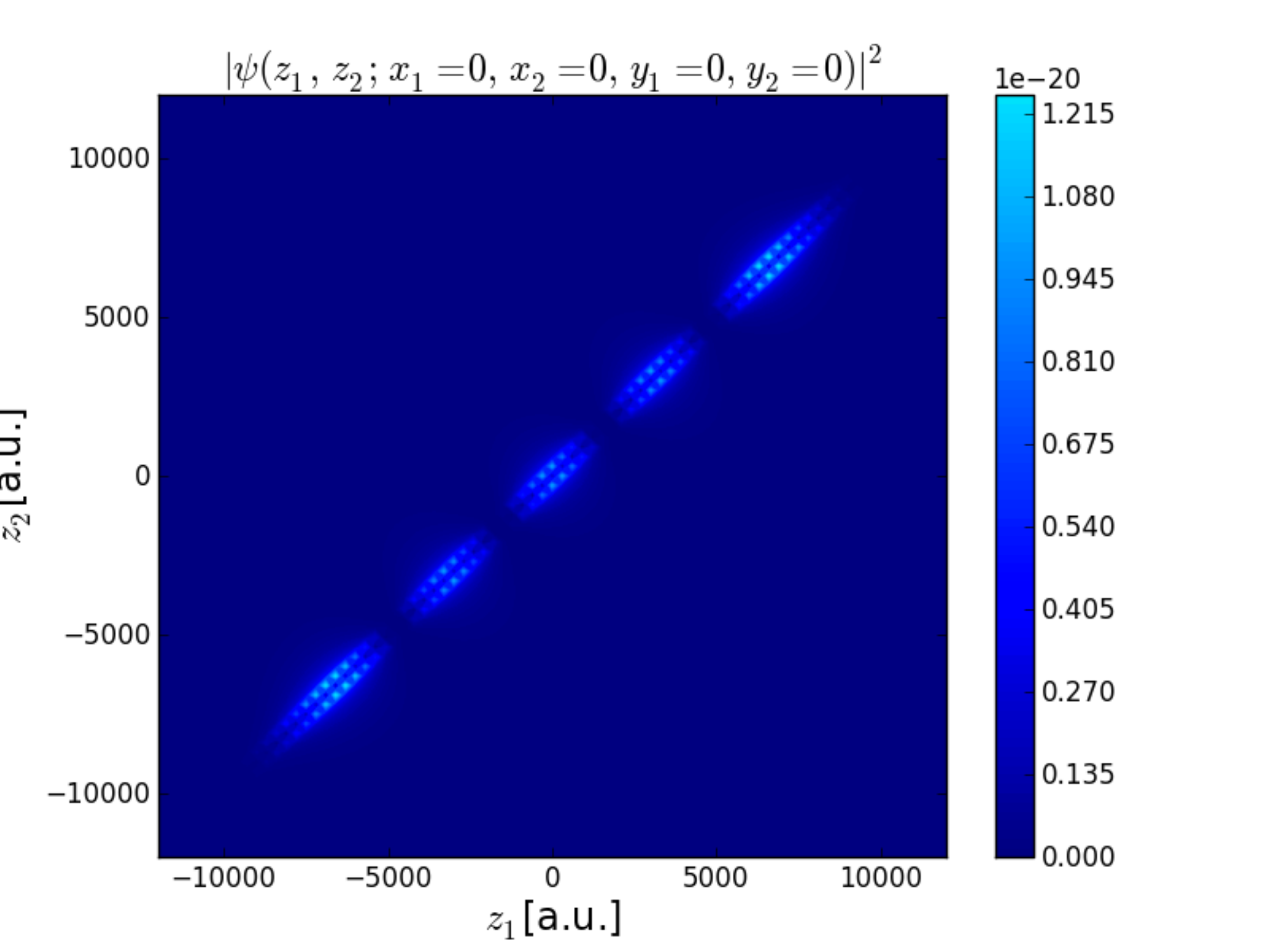}
  \caption{(Color online) Cuts along the elongated z-direction
    ($\left|\Psi(z_1,z_2;x_1=x_2=y_1=y_2=0)\right|^2$) through the full
    six-dimensional \textit{ab initio} wavefunction.\\
    Upper part: The c.m.\ excited bound state diabatically described by
    $\ket{\psi_b \Phi_{(2,0,0)}}$ at $d_y/a
    = 1.42$.\\
    Lower part: The c.m.\ excited bound state diabatically described by
    $\ket{\psi_b \Phi_{(0,0,4)}}$ at $d_y/a = -1.2$.\\
    (For both plots an identical color code has been used.)}
  \label{fig:cntr_bound}
  \end{centering}
\end{figure}

\begin{figure}[ht]
  \begin{centering}
  \includegraphics[width=0.44\textwidth]{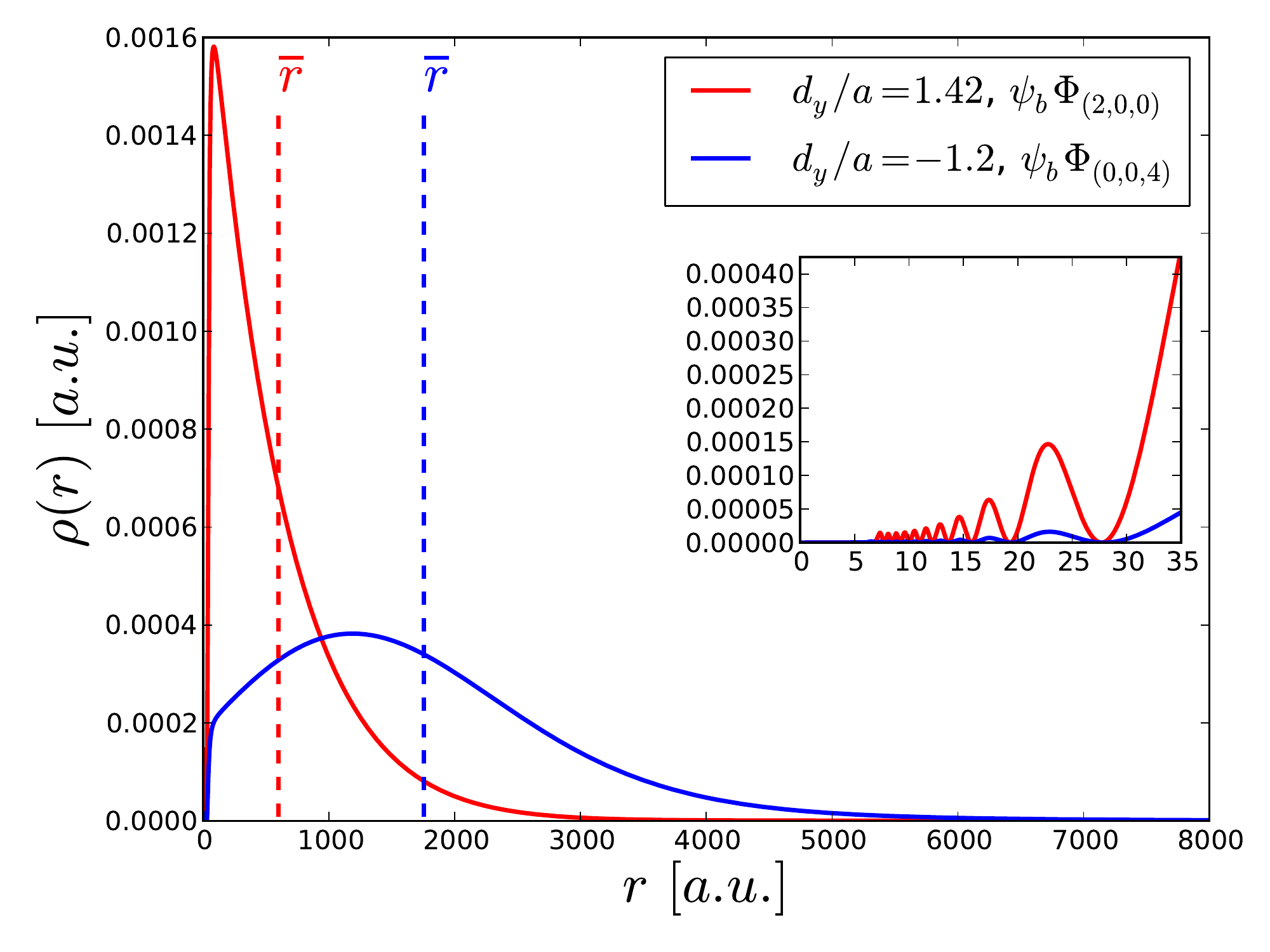}
  \caption{(Color online) Radial density of the c.m.\ excited bound
    states.}
  \label{fig:raddens_-1d3}
  \end{centering}
\end{figure}

In \figref{fig:cntr_bound} cuts through the bound-state densities are
shown. In both states, the particles only occupy regions where they
are very close to each other, i.\,e.\ close to the diagonal
$z_1=z_2$. The bound state at $d_y/a = 1.42$ (upper panel in
\figref{fig:cntr_bound}) has no c.m.\ excitation in the $z$
direction. Hence, no zeros (nodes) are visible in the density
(wavefunction) in scales of the trap length. Of course, the many
small-scale oscillations in the bound-state regime stemming from the
deep Born-Oppenheimer interaction potential are still present. The
bound state at $d_y/a=-1.2$ shows four large-scale nodes along the
$z$ direction which is due to the $(0,0,4)$ c.m.\ excitation of this
bound state.

At the transversally excited resonance, at $d_y/a = 1.42$, see
\figref{fig:spec}, the atoms in the bound state which can be
approximated by $\ket{\psi_{(b)}\, \Phi_{(2,0,0)}}$ have a mean
distance of $\overline{r_{b}} = 0.29\, d_{y}$, i.\,e.\ it is small
compared to the confinement length in the tight $y$ direction. This
demonstrates the strong binding of the atoms. Away from the $(0,0,4)$
resonance, at $d_y/a=-1.2$, the atoms in the bound state have a mean
distance of $\overline{r_{b}} = 1.01\, d_{y} = 92.9\,$nm, i.\,e.\ it is
on the order of the trap length in the tightly confined direction.

An interesting question is whether the bound state at the $(0,0,4)$
resonance at $d_y/a=-1.2$ has enough bound character to trigger 
molecule formation and subsequent losses in a many-body system.

In \cite{cold:sala13} a molecule formation was observed experimentally
at the resonance where the atoms in the bound state had a mean radial
distance of $\overline{r_b} = 140\,$nm. This is even larger than the
value of $\overline{r_b}=96.8\,$nm at the $(0,0,4)$ resonance. Hence,
a molecule formation with subsequent processes is also expected at
this resonance.

\section{Conclusion}
\label{sec:conclusion}

Experiments \cite{cold:hall10b,cold:sala13} have demonstrated that
inelastic confinement induced-resonances can influence the stability 
of an ultracold quantum gas
and can be adopted to create molecules fully coherently. The
resonances detected so far were measured in a strongly anisotropic
confinement at large positive values of the $s$-wave scattering
length. In fact, it has been demonstrated that they were all of a
special kind, namely the ones excited in the strongly confined
direction. In the present work it is demonstrated that also resonances
in the weakly confined direction occur. Models are introduced to
describe the resonance position and the coupling strength of, in
principle, all c.m.-rel.\ coupling resonances of a system of two
ultracold atoms.

A study of the most pronounced resonances is performed for a variation
of the external confinement. The lowest-order resonance in the
strongly confined direction(s) and the next to leading order
resonances in the weakly confined direction(s) are discussed in the
transition of an almost isotropic 3D confinement to a quasi-1D 
(cigar-shaped) and a quasi-2D (pancake-shaped) confinement. While the
position and the coupling strength of the resonance(s) excited in the
strongly confined direction converge monotonically to a constant
(non-zero) value for an increasing anisotropy, the position of the
resonance(s) excited in the weakly confined direction fade away to
negative infinity for an increasing anisotropy and the coupling
strength approaches zero. These resonances show a maximum in the
coupling strength for intermediate anisotropies.

The models are discussed in comparison to \textit{ab initio}
calculations. In the transition to a cigar-shaped confinement
geometry, the resonance positions are described accurately by the
model C. The coupling strength is described quantitatively correct by
model 1 for the resonances in the strongly confined direction and
qualitatively correct for the resonance in the weakly confined
direction.

In the transition to a pancake-shaped geometry, the resonance position
of the resonance with an excitation in the strongly confined direction
is described quantitatively accurate. For the resonances in the weakly
confined direction, the accuracy for different anisotropies depends on
the used approximations. The coupling strengths are reproduced
qualitatively by the model 2 except for large anisotropies. For the
latter the model 2 faces limitations and results in negative coupling
strengths.

A variation of the potential depth shows that the c.m.-rel.\ coupling
can be increased by decreasing the potential depth. While the position
of the resonances in a very shallow sextic potential is still
described accurately by the model B, the values for the coupling
strengths loose their accuracy, since the harmonic approximation of
the wavefunctions the model is based on looses its validity.
Therefore, while the discussed models provide a helpful guide, the
full \textit{ab initio} calculation remains indispensable for highly
precise quantitative predictions or for describing properly some trap
geometries.

The analysis of the wavefunctions involved in the resonances in a
cigar-shaped potential demonstrates that molecule formation and
subsequent losses are also expected for the resonance excited in the
weakly confined direction. In this case they occur for large negative values
of the $s$-wave scattering length. We hope that this type of resonance will
soon be verified experimentally.

The study of inelastic CIR has demonstrated that one of the most
fundamental and routinely adopted approximations in ultracold atomic
quantum gases --- the harmonic approximation --- has to be abandoned
in order to describe particle loss, heating, and molecule formation in
a variety of experiments. In fact, the inelastic CIR can be tuned not
only by a variation of the scattering length but alternatively by a
modification of the confinement geometry. Hence, it might deliver a
novel tool for ultracold-atom experiments to alter the interaction
behavior by a variation of the external confinement in the vicinity of
an inelastic CIR. This can be valuable in cases where the standard
technique of using magnetic Feshbach resonances may not be available,
such as for earth-alkali atoms. Similar to a magnetic Feshbach
resonance, at the inelastic CIR the association of molecules can be
tuned fully coherently.

\acknowledgments{The authors gratefully acknowledge financial support
  from the \textit{Studienstiftung des deutschen Volkes} and the
  \textit{Fonds der Chemischen Industrie}.}
%

%

\section{Appendix}
\subsection{2-Channel model}
\label{sec:2channel}

To obtain the resonance position and coupling strength from \textit{ab
  initio} calculations, a two-channel model is fitted to the
eigenenergy spectrum in the vicinity of the investigated
resonance. There are two diabatic states, the trap state $\ket{t}$ and
the bound state $\ket{b}$ with diabatic energies $E_t$ and $E_b$. 
Introducing a coupling $W$ between these states, the Hamiltonian matrix 
\begin{align}
  H =
  \begin{pmatrix}
    E_t & W \\
    W  & E_b 
  \end{pmatrix}
\end{align}
is obtained. 
A diagonalization of this matrix by a linear transformation
$H_{\mathrm{d}} = U^{-1}HU$, where U consists of the eigenvectors of
the diagonal matrix $H_{\mathrm{d}}$, leads to the energies $E_1$ and
$E_2$ of the adiabatic states which are known from the \textit{ab
  initio} calculations. Assuming that the diabatic states are linear
in the vicinity of the avoided crossing, i.\,e.\ $E_t = ax +b$ and
$E_b=cx +d$ (where $x=\frac{d_y}{a}$), the coefficients $a,b,c,d$ and
the coupling $W$ can be obtained by a minimization $\left\Vert
  U^{-1}HU - H_{\mathrm{d}}\right\Vert$. The position of the resonance
is then easily obtained from the crossing point of $E_t(x)$ and
$E_b(x)$.

\subsection{Perturbation Theory }
\label{sec:perturb}

For a correction of the resonance positions in the model B, the
energies of the c.m.\ trap states in the sextic potential are treated
within first-order perturbation theory. Since the c.m.\ Hamiltonian
separates, it is sufficient to evaluate the 1D Hamiltonians. The
unperturbed system is the 1D harmonic oscillator for which the wavefunctions
are given in \eqref{eq:ho_1d}. The anharmonic terms of the sextic
potential
\begin{align}
  V^{(\mathrm{a})}_j(R_j)= -\frac{1}{24}\frac{\hbar \omega_j}{V_j}R_j^4
  + \frac{1}{720}\frac{\hbar^2 \omega_j^2}{V_j^2}R_j^6
\end{align}
are treated as a perturbation. Here the potential is written in
dimensionless units of energies in $\hbar \omega_j$ and lengths in
$\sqrt{\frac{\hbar}{M \omega_j}}$, where $M=2m$. For simplicity, only a
single spatial direction is considered in the following and the
subscript $j$ is omitted.
The first-order energy correction is determined by
\begin{align}
  \label{eq:energy_correc}
  E_n^{(\mathrm{a})}= \int_{-\infty}^{\infty} dR\,  |\psi(R)|^2 V^{(\mathrm{a})}(R).
\end{align}
An exact expression for the integral of a triple product of Hermite
polynomials and a Gaussian is known to be \cite{cold:grad07} 
\begin{align}
 \label{eq:grad_Herm}
 \int_{-\infty}^{\infty} dx\, \mathrm{e}^{-x^2} \mathrm{H}_k(x)
 \mathrm{H}_n(x) \mathrm{H}_m(x)= \frac{2^{\frac{m+n+k}{2}} \sqrt{\pi}
   k! n! m! }{ (s-k)!(s-n)!(s-m)!}
\end{align}
where $2s=n+k+m$ must be even. To make use of this formula the $R^4$
and $R^6$ terms in $V^{(\mathrm{a})}$ need to be expressed in Hermite
polynomials. For example,
\begin{align}
  R^4=\frac{1}{16} \mathrm{H}_4(R) + \frac{3}{4} \mathrm{H}_2(R) +
  \frac{3}{4} \mathrm{H}_0(R).
\end{align}
Inserting the expressions for $R^4$ and $R^6$ into the integral in 
\eqref{eq:energy_correc}, splitting the integrals, and evaluating each
with formula \eqref{eq:grad_Herm} yields
\begin{align}
  E_n^{(\mathrm{a})} = &-\frac{1}{1152 \, V^{2}}\ \bigg[ 36 \, {\left(2
        \, n^{2} + 2 \, n + 1\right)} V \hbar^2 \omega^{2} \nonumber\\ 
    &-{\left(4 \, n^{3} + 6 \, n^{2} + 8 \, n + 3\right)} \hbar^3
    \omega^{3} \bigg] \quad .
\end{align}
To determine the 3D perturbative energies $E_n =\sum_{j=x,y,z}
E_{n,j}^{(\mathrm{h})} + E_{n,j}^{(\mathrm{a})}$ of the sextic potential
the anharmonic energy corrections $E_{n,j}^{(\mathrm{a})}$ of the
three spatial directions and the corresponding harmonic oscillator
energies $E_n^{(\mathrm{h})} = \hbar \omega_j\, (n+\frac{1}{2})$ need
to be added up.

\end{document}